\documentstyle[aps,prbbib,epsfig,floats]{revtex}
\begin{document}
\wideabs{
\title{Exchange and spin-fluctuation superconducting pairing
in cuprates}
\author{N. M. Plakida$\sp{a}$, L. Anton$\sp{b,a,d}$, S. Adam$\sp{c,a}$,
Gh. Adam$\sp{c,a}$}
\address{
$\sp{a}$Joint Institute for Nuclear Research, 141980 Dubna, Moscow, Russia\\
$\sp{b}$Institute of Atomic Physics INFLPR, Lab.22 PO Box
MG-36, R-76900, Bucharest, Romania\\
$\sp{c}$Institute of Physics and Nuclear Engineering, Department
of Theoretical Physics, PO Box MG-6, R-76900,
Bucharest-M{\v{a}}gurele, Romania,\\
$\sp{d}$Institute for Theoretical Physics, University of Stellenbosch,
Private Bag X1, 7602 Matieland, South Africa}
\date{\today}
\maketitle
\begin{abstract}
We propose a microscopical theory of superconductivity in CuO$_2$
layer within the effective two-band  Hubbard model in the strong
correlation limit.  By applying a projection technique for the
matrix Green function in terms of the Hubbard operators, the Dyson
equation is derived. It is proved that in the mean-field
approximation $d$-wave superconducting pairing mediated by the
conventional exchange interaction occurs.  Allowing for the
self-energy corrections due to kinematic interaction,
a spin-fluctuation $d$-wave   pairing is also obtained.
$T\sb{c}$ dependence on the hole concentration and
$\bf k$-dependence of the gap function are derived. The results show
that the exchange interaction (which stems from the interband
hopping) prevails over the kinematic interaction (which stems
from the intraband hopping).
\end{abstract}
}
\section{Introduction}
\label{sec:intro}

Recent experiment on field-induced superconductivity in infinite
layer cuprate compound CaCuO$_2$ has proved that superconducting
pairing is confined to the copper-oxygen plane both for electron
and hole doping.\cite{Schon01} Therefore to describe the mechanism
of superconducting pairing in cuprates  one should study
interactions in one copper-oxygen plane. As was initially pointed
out by Anderson~\cite{Anderson87}, strong electron correlations in
copper $3d$-states play the most important role in explaining
antiferromagnetic (AFM) and insulating properties of the undoped
plane. These correlations in the electronic liquid of doped
electrons or holes could be also responsible for superconducting
pairing in the plane.~\cite{Scalapino95}
\par
Electron correlations are usually considered within the framework
of  the Hubbard model or the so-called $t$-$J$ model which can be
derived from the Hubbard model in the limit of strong
correlations.~\cite{Anderson87} Superconducting pairing mediated
by AFM spin-fluctuations in the weak correlation limit of the
Hubbard model has been proved  by using different approaches: by
applying AFM spin-wave model~\cite{Kampf90}, within the
fluctuation exchange approximation (FLEX) (see,
e.g.~\cite{Pao1994,Manske00} and the references therein), a
renormalization group technique (see~\cite{Halboth00,Irkhin01} and
the references therein).  A number of phenomenological models like
the nearly AFM Fermi liquid\cite{Monthoux93} was also proposed.
\par
 In the strong correlation limit a lot of numerical work has been
done~\cite{Scalapino95,Dagotto94} though the obtained results are
still controversial. For instance, a robust $d$-wave pairing was
observed for the $t$-$J$ model,~\cite{Sorella01} while absence of
the long-range order was reported for the original Hubbard
model.~\cite{Huang01} Concerning analytical calculations, usually
a mean-field approximation (MFA)  both for the $t$-$J$ model (see,
e.g.,~\cite{Plakida89}) and for the Hubbard
model~\cite{Plakida88,Beenen95,Avella97,Stanescu00} were employed.
Self-energy corrections were considered only for the $t$-$J$ model
within an involved  diagram technique~\cite{Izyumov92} or by using
the Mori-type projection technique for the Green
functions~\cite{Plakida99}. A numerical solution of the Dyson
equation in~\cite{Plakida99} revealed a non Fermi-liquid behavior
 in the normal state, while  the $d$-wave
superconductivity mediated by the exchange and spin-fluctuation
pairing was observed. To clear up the origin of the controversial
numerical results reported for the $t$-$J$  and Hubbard models and
to elucidate the pairing mechanism in cuprates a comprehensive
study of the Hubbard, or more general $p$-$d$ model beyond the MFA
 is very important.
\par
In the present paper, we consider a microscopical theory of
superconductivity in CuO$_2$ layer within the effective two-band
Hubbard model~\cite{Plakida95a} in the strong correlation limit.
We have proved, by a direct calculation of  anomalous correlation
functions in MFA     that  the $d$-wave pairing  in the model is
mediated by the conventional AFM exchange interaction as in the
$t$-$J$ model. The retardation effects in the exchange
interaction, which originate from the interband hopping  with a
large excitation energy, are negligible that results in the
pairing of all the electrons (holes) in a conduction subband and
a high $T_c$ proportional to the Fermi energy. Allowing for the
self-energy corrections beyond the MFA, we consider also the
spin-fluctuation pairing induced by kinematic interaction in the
second order. The latter, acting in a narrow energy shell of the
order of a spin-fluctuation energy $\omega_s$, results in a lower
$T_c$.
 \par
 The paper is organized as follows. In Sec.~\ref{sec:geral}, the
general formalism and the Dyson equation for the matrix GF in
Nambu notation are derived for the two-band Hubbard model.
Sec.~\ref{sec:m-f} is devoted to the discussion of the
superconducting pairing within MFA. The self-energy is calculated
in Sec.~\ref{sec:s-e} within the self-consistent Born
approximation. Numerical results and their discussion are provided
in Sec.~\ref{sec:numeric}. Conclusions are presented in
Sec.~\ref{sec:concl}.

\section{General formalism}
\label{sec:geral}

A distinctive property of cuprates is  the strong
antiferromagnetic superexchange interaction for copper spins that
reaches a record value of 1500~K for transition metal compounds.
It is caused by the strong $pd\sigma$ hybridization $t_{pd}
\simeq 1.5$~eV  for the  $3d$-copper states and  the $2p$-oxygen
states and the small  splitting energy of their atomic levels
$\Delta_{pd} \simeq 3$~eV.  At the same time strong Coulomb
correlations for the $3d$ states of copper, $U_d \simeq 8$~eV,
considerably increase the energy of two-hole $3d$ states and the
lowest energy level at hole doping appears to be a Zhang-Rice
singlet~\cite{Zhang88}. These features of the electronic spectrum
in CuO$_2$ plane can be described within the framework of the
$p$-$d$   model Hamiltonian~\cite{Emery87}:
\begin{eqnarray}
  H & = &  \sum_{i \sigma} \{
 \epsilon_d \ \tilde{d}_{i \sigma}^+\tilde{d}_{i \sigma}
+ \epsilon_p \ c_{i \sigma}^+ c_{i \sigma} \}
\nonumber\\
 & + & \sum_{i, j, \sigma} V_{i j} \ \{ \tilde{d}_{i \sigma}^+
 c_{j\sigma} + {\rm H.c.} \}.
\label{eq:pd}
\end{eqnarray}
Operators $\tilde{d}_{i \sigma}^+ $ and $\ c_{i \sigma}^+ $ describe the
creation of one-hole $d$ and  $p$ states at sites $i$ of the square
lattice in the  CuO$_2$ plane with the energies
 $\epsilon_d$ and
$\epsilon_p = \epsilon_d + \Delta_{pd}$, respectively. Because of
the large Coulomb correlations,  only singly occupied $3d$ states
are taken into account:  $\tilde{d}_{i \sigma}^+ = {d}_{i
\sigma}^+ (1- n^{d}_{i, -\sigma})$. For the bonding oxygen
orbitals the Wannier representation is used that  results in the
$p$-$d$ hybridization parameters $\, V_{i j} = 2 t_{pd} \nu_{ij}
\,$ with the coefficients~\cite{Plakida95a} $\, \nu_0=\nu_{j j}
\simeq 0.96, \; \nu_1 = \nu_{j \ j\pm a_{x/y}} \simeq -0.14, \;
\nu_2 = \nu_{j \ j \pm a_x \pm a_y} \simeq -0.02 \, $ for the
single-site hybridization $\nu_0$ and the  hybridization between
the nearest $\nu_1$ and the next  nearest $\nu_2$ neighbor sites,
respectively, where $a\sb{x/y}$ are the lattice constants. Since
the single-site hybridization  is quite large, $\, V_0 \simeq
 \Delta_{pd} \,$,  and $ V_0 \gg |V_{i \neq j}| \,$ one has
 at first to diagonalize
the single-site part  of the Hamiltonian~(\ref{eq:pd}) and then
to apply a perturbation theory for intersite hopping. As a result
of this cell-perturbation theory (see,
e.g.,~\cite{Plakida95a,Feiner96,Yushankhai97}), one gets an
effective Hubbard model with the lower Hubbard subband occupied
by one-hole Cu-$d$-like states and the upper Hubbard subband
occupied by two-hole $p$-$d$-like  singlet states
\begin{eqnarray}
  H&=&E\sb{1} \sum\sb{i,\sigma} X\sb{i}\sp{\sigma \sigma} +
      E\sb{2} \sum\sb{i} X\sb{i}\sp{22} + \sum\sb{i\neq j,\sigma}
    \bigl\{
      t\sb{ij}\sp{11} X\sb{i}\sp{\sigma 0} X\sb{j}\sp{0\sigma}
\nonumber\\
   &+&t\sb{ij}\sp{22} X\sb{i}\sp{2 \sigma} X\sb{j}\sp{\sigma 2} +
      2\sigma t\sb{ij}\sp{12} (X\sb{i}\sp{2\bar\sigma} X\sb{j}\sp{0 \sigma} +
      {\rm H.c.})
    \bigr\} ,
\label{eq:H}
\end{eqnarray}
where we introduced the Hubbard operators: $X\sb{i}\sp{nm} =
|in\rangle\langle im|$ for the four states $n,m=|0\rangle
,\,|\sigma\rangle ,\, |2\rangle =|\uparrow \downarrow \rangle $,
$\sigma=\pm 1/2$, $\bar\sigma=-\sigma$. The energy parameters are
given by $E\sb{1}=\tilde{\epsilon}\sb{d}-\mu$ and
$E\sb{2}=2E\sb{1}+\Delta$ respectively, where
$\tilde{\epsilon}\sb{d}$ is a reference (renormalized) energy of
the $d$-hole, $\mu$ is the chemical potential, and $\Delta \simeq
\Delta_{pd}$ is the renormalized charge transfer energy
(see~\cite{Plakida95a}). Here and in what follows, the
superscripts $1$ and $2$ refer to the one-hole and singlet
subbands, respectively. The hopping parameters can be written as
$t\sp{\alpha\beta}\sb{ij} =
 K\sb{\alpha\beta}\, V\sb{ij} \;$.
 The coefficients
$K\sb{\alpha\beta}$ depend on the dimensionless parameter
$t_{pd}/{\Delta_{pd}}$. Assuming the realistic value $\Delta_{pd}
= 2t_{pd}$, we get for these  parameters~\cite{Plakida95a}:
$K\sb{11}\simeq -0.89$, $\, K\sb{22}\simeq -0.48$, $\, K\sb{12}
\simeq 0.83$. Therefore, the effective hopping parameters for the
nearest-neighbors in the model~(\ref{eq:H}) are given by
$t\sb{eff} \simeq | K\sb{22}|\, 2\nu\sb{1}\, t_{pd} \simeq 0.14
t_{pd} \,$ that results in a narrow bandwidth $W = 8 t\sb{eff}
\simeq t_{pd} $. Since in~(\ref{eq:H}) the charge-transfer gap
$\Delta$ plays the role of the Coulomb repulsion $U$ in the
conventional Hubbard model, the effective Hubbard
model~(\ref{eq:H}) corresponds to the strong correlation limit
due to a large ratio of $\Delta$ to the band width: $\, \Delta /
W \simeq 2 \,$. Therefore at half-filling the
Hamiltonian~(\ref{eq:H}) describes the Mott-Hubbard insulating
state (see Fig.1a in Ref.~\cite{Plakida95a}).
\par
In the present paper we have assumed a simple version of the
$p$-$d$ model~\cite{Emery87} which has only two fitting
parameters, $t_{pd}$ and $\Delta_{pd}$, in Eq.~(\ref{eq:pd}).
Starting from a more general model which takes into account
finite values of the Coulomb repulsion on $d$-sites $U_d$,   on
$p$-sites $U_p$, and between the nearest neighbor $p$-$d$ sites
$U_{pd}$,  and $p$-$p$ hybridization $t_{pp}$ one arrives to the
same effective Hubbard model~(\ref{eq:H}) but with renormalized
hopping parameters $t\sp{\alpha\beta}\sb{ij}$ and charge transfer
energy $\Delta$ (see, e. g.,~\cite{Feiner96,Yushankhai97}).
\par
The Hubbard operators entering~(\ref{eq:H}) obey the
multiplication rules: $\, X\sb{i}\sp{nm}X\sb{i}\sp{kl} =
\delta_{m,k} X\sb{i}\sp{nl} \,$ and  the completeness relation
\begin{equation}
  X\sb{i}\sp{00} + X\sb{i}\sp{\sigma\sigma} +
  X\sb{i}\sp{\bar\sigma\bar\sigma} + X\sb{i}\sp{22} = 1.
\label{eq:no2oc}
\end{equation}
The latter rigorously preserves the constraint of no double
occupation at each lattice site $i$  by any quantum state
$\,|in\rangle.$
\par
To discuss the quasi-particle (QP) spectrum and superconducting
pairing within the model Hamiltonian~(\ref{eq:H}), we introduce
the two-time anticommutator retarded $4\times 4$ matrix Green
function (GF)  in Zubarev notation~\cite{Zubarev60}
\begin{eqnarray}
& &  \tilde G\sb{ij\sigma}(t-t')  =
    \langle\langle \hat X\sb{i\sigma}(t)\! \mid \!
    \hat X\sb{j\sigma}\sp{\dagger}(t')\rangle\rangle
 \nonumber\\
&=& \int\limits_{-\infty}^{+\infty}
\frac{{\rm d}\omega} {2\pi}\; {\rm e}^{-i\omega (t-t^{\prime})}\,
\frac{1}{N}\sum_{\bf q} {\rm e}^{i{\bf q\cdot (i - j)}} \;
\tilde G_{\sigma}({\bf q}, \omega),
\label{eq:GF}
\end{eqnarray}
for  the  four-component Nambu operators~$\hat
X\sb{i\sigma}\sp{\dagger}$
\begin{equation}
  \hat X\sb{i\sigma}\sp{\dagger}=(X\sb{i}\sp{2\sigma}\,\,
  X\sb{i}\sp{\bar\sigma 0}\,\, X\sb{i}\sp{\bar\sigma 2}\,\,
  X\sb{i}\sp{0\sigma})  ,
\label{eq:nambucr}
\end{equation}
and $\hat X\sb{i\sigma}$  obtained from~(\ref{eq:nambucr}) by
Hermitian conjugation.
 The GF~(\ref{eq:GF}) can thus be written as a $2\times 2$
supermatrix of normal, $\hat G\sb{ij\sigma}(\omega)$, and
anomalous, $\hat F\sb{ij\sigma}(\omega)$, $2\times 2$ matrix
components:
\begin{equation}
  \tilde G\sb{ij\sigma}(\omega) =
  \left( \begin{array}{cc}
    \hat G\sb{ij\sigma}(\omega)& \hat F\sb{ij\sigma}(\omega)\\
    \hat F\sb{ij\sigma}\sp{\dagger}(\omega) &
      - \hat {G}\sb{ij\bar\sigma}\sp{({\rm T})}(-\omega)
  \end{array} \right) \, ,
\label{eq:gf2x2}
\end{equation}
where the superscript (T) denotes the operation of transposition.
\par
To calculate the GF~(\ref{eq:GF})  we use the equation of motion
method. Differentiation with respect to time $t$ of the GF
(\ref{eq:GF}) and use of the Fourier transform  result in the
following equation \cite{Plakida97}
\begin{equation}
  \omega \tilde G\sb{ij\sigma}(\omega) = \delta \sb{ij} \tilde \chi +
   \langle\!\langle
    \hat Z\sb{i\sigma} \!\mid\! \hat X\sb{j\sigma}\sp{\dagger}
   \rangle\!\rangle\sb{\omega}\, ,
\label{eq:GFom}
\end{equation}
where $\hat Z\sb{i\sigma}=[\hat X\sb{i\sigma},H]$.  Assuming that
the system is in the paramagnetic state, $\,\langle
X\sb{i}\sp{\sigma\sigma} \rangle =
   \langle  X\sb{i}\sp{\bar\sigma\bar\sigma} \rangle \,$,
we get for the matrix $\, \tilde\chi=\langle \{\hat
X\sb{i\sigma},\hat X\sb{i\sigma}\sp{\dagger}\}\rangle $:
\begin{equation}
  \tilde\chi= \left(
    \begin{array}{cccc}
      \chi\sb{2} & 0 & 0 & \chi\sb{3}\\
      0 & \chi\sb{1} & \chi\sb{3} & 0\\
      0 &\chi\sb{3}\sp{*}& \chi\sb{2} & 0\\
      \chi\sb{3}\sp{*} & 0& 0 & \chi\sb{1}
    \end{array}
              \right).
\label{eq:chicalc}
\end{equation}
For the diagonal terms of the matrix we get   $\chi\sb{2} =
\langle X\sb{i}\sp{22} + X\sb{i}\sp{\sigma\sigma} \rangle
 =  n/2$, $\chi\sb{1} = \langle X\sb{i}\sp{00} +
X\sb{i}\sp{\bar\sigma \bar\sigma} \rangle = 1-\chi\sb{2}$
 where $n = 1+ \delta$ denotes the hole concentration. The
anomalous correlation function
 $\, \chi\sb{3} = \langle X\sb{i}\sp{02}\rangle \, $ describes
single-site pairing  and for the $d$-wave symmetry it vanishes
\begin{equation}
  \chi\sb{3} = \langle X\sb{i}\sp{02} \rangle =
 \langle c\sb{i\downarrow}c\sb{i\uparrow}\rangle = 0.
\label{chi3}
\end{equation}
Here we used  the identity
\begin{equation}
  X\sb{i}\sp{02} = X\sb{i}\sp{0\downarrow} X\sb{i}\sp{\downarrow 2} =
   c\sb{i\downarrow}c\sb{i\uparrow} \, ,
\label{eq:x2da}
\end{equation}
which results from the definitions of the Fermi annihilation
operators: $ c\sb{i\sigma} = X\sb{i}\sp{0\sigma} + 2\sigma
X\sb{i}\sp{\bar\sigma 2}$ and the multiplication rules  for the
Hubbard operators,
 $X\sb{i}\sp{0\sigma} X\sb{i}\sp{0 \bar\sigma} = 0$,
 $X\sb{i}\sp{\sigma 2} X\sb{i}\sp{\bar\sigma 2} = 0$,
 $X\sb{i}\sp{0\downarrow} X\sb{i}\sp{\downarrow 2} = X\sb{i}\sp{02}$.
\par
The chemical potential $\mu$ is calculated from the equation for the
average number of holes,
\begin{equation}
  n = \langle N\sb{i} \rangle
    = \sum\sb{\sigma} \langle X\sb{i}\sp{\sigma \sigma} \rangle +
      2 \langle X\sb{i}\sp{22} \rangle \; .
\label{def-n}
\end{equation}
\par
Now by using the Mori-type projection technique we write the equation of
motion for the operator $\hat Z\sb{i\sigma}$ in Eq.~(\ref{eq:GFom}) as
a sum of a linear part  and an irreducible part orthogonal to it,  $\hat
Z\sb{i\sigma}\sp{(ir)}$, which originates from the inelastic QP
scattering:
\begin{equation}
  \hat Z\sb{i\sigma} = [\hat X\sb{i\sigma}, H] =
    \sum\sb{l}\tilde E\sb{il\sigma} \hat X\sb{l\sigma} +
    \hat Z\sb{i\sigma}\sp{(ir)}.
\label{eq:irred}
\end{equation}
 The orthogonality condition
 $\, \langle \{ \hat Z\sb{i\sigma}\sp{(ir)},
    \hat X\sb{j\sigma}\sp{\dagger} \} \rangle = 0 \,$
provides the definition of the  the frequency matrix:
\begin{eqnarray}
 \tilde E\sb{ij\sigma}& = &\tilde {\cal A}\sb{ij\sigma} \tilde\chi\sp{-1},
\label{eq:freq}\\
  \tilde {\cal A}\sb{ij\sigma}& =& \langle \{ [\hat X\sb{i\sigma}, H],
    \hat X\sb{j\sigma}\sp{\dagger} \} \rangle.
\label{eq:Aij}
\end{eqnarray}
The frequency matrix~(\ref{eq:freq}) defines the zero-order GF
in  the generalized MFA. In the
$({\bf q}, \omega )$-representation, its expression is given by
\begin{equation}
  \tilde G\sp{0}\sb{\sigma }({\bf q},\omega) =
    \Bigl( \omega \tilde \tau\sb{0} - \tilde E\sb{\sigma}({\bf q})
      \Bigr) \sp{-1} \tilde \chi\;,
\label{eq:gf0}
\end{equation}
where $\tilde \tau\sb{0}$ is the $4\times 4$ unity matrix.
\par
Differentiation of the many-particle GF~(\ref{eq:GFom}) with respect
to the second time $t'$ and use of the same projection procedure as
in~(\ref{eq:irred}) result in the Dyson equation for the
GF~(\ref{eq:GF}). In $({\bf q}, \omega )$-representation, the
Dyson equation is
\begin{equation}
  \left( \tilde G\sb{\sigma}({\bf q}, \omega) \right)\sp{-1} =
  \left( \tilde G\sb{\sigma}\sp{0}({\bf q}, \omega) \right)\sp{-1} -
  \tilde \Sigma\sb{\sigma}({\bf q}, \omega).
\label{eq:Dyson}
\end{equation}
The self-energy operator $\tilde\Sigma\sb{\sigma}({\bf q}, \omega)$ is
defined by
\begin{equation}
  \tilde T\sb{\sigma}({\bf q}, \omega) =
  \tilde \Sigma\sb{\sigma}({\bf q}, \omega) +
  \tilde \Sigma\sb{\sigma}({\bf q}, \omega)
  \tilde G\sp{0}\sb{\sigma}({\bf q}, \omega)
  \tilde T\sb{\sigma}({\bf q}, \omega),
\label{eq:selfenergy}
\end{equation}
where
$\;  \tilde T\sb{\sigma}({\bf q}, \omega) = {\tilde \chi}\sp{-1}
   \langle\!\langle
     {\hat Z}\sb{{\bf q}\sigma}\sp{(ir)} \!\mid\!
     {\hat Z}\sb{{\bf q}\sigma}\sp{(ir)\dagger}
   \rangle\!\rangle\sb{\omega}
  {\tilde \chi}\sp{-1} \;$
denotes the scattering matrix. From Eq.~(\ref{eq:selfenergy}) it follows
that the self-energy operator is given by the {\it proper\/} part of the
scattering matrix that has no parts connected by the
single-particle zero-order GF~(\ref{eq:gf0}):
\begin{equation}
  \tilde \Sigma\sb{\sigma}({\bf q}, \omega) = {\tilde \chi}\sp{-1}
    \langle\!\langle {\hat Z}\sb{{\bf q}\sigma}\sp{(ir)} \!\mid\!
     {\hat Z}\sb{{\bf q}\sigma}\sp{(ir)\dagger} \rangle\!\rangle
      \sp{(prop)}\sb{\omega}\;{\tilde \chi}\sp{-1} .
\label{eq:self-enir-qo}
\end{equation}
\par
The equations~(\ref{eq:gf0}),~(\ref{eq:Dyson}),
and~(\ref{eq:self-enir-qo}) provide an exact representation for
the single-particle GF~(\ref{eq:GF}). Its calculation, however,
requires the use of some approximations for the many-particle GF
in the self-energy matrix~(\ref{eq:self-enir-qo}) which describes
the finite lifetime effects (inelastic scattering of electrons on
spin and charge fluctuations).

\section{Mean-field approximation}
\label{sec:m-f}
\subsection{Spectrum in the normal state}
\label{sec:normal}
Let us consider at first the QP spectrum in the  generalized MFA
 given by the  zero-order GF~(\ref{eq:gf0}). The latter is
defined by the frequency matrix~(\ref{eq:freq}) which can be
readily calculated by writing down equations of motion for the
Hubbard operators in~ Eq.~(\ref{eq:nambucr}) and performing
necessary commutations in the matrix $\tilde {\cal
A}\sb{ij\sigma}$,~Eq.~(\ref{eq:Aij}). As a result of these
calculations, which are  presented at greater length in the
Appendix A,  the matrix~(\ref{eq:Aij})  can be written in the form
\begin{equation}
  \tilde {\cal A}\sb{ij\sigma} = \left(
    \begin{array}{cc}
      \hat{\omega}\sb{ij\sigma} & \hat{\Delta}\sb{ij\sigma}\\
      (\hat{\Delta}\sb{ij\sigma}\sp{*})\sp{(\rm{T})} &
         -\hat{\omega}\sb{ij\bar\sigma}\sp{(\rm{T})}
    \end{array} \right),
\label{aij}
\end{equation}
where $\hat{\omega}\sb{ij\sigma}$ and $\hat{\Delta}\sb{ij\sigma}$
are $\, 2\times 2 \, $ matrices. The diagonal parts of the matrix
defines the normal state QP spectrum:
\begin{eqnarray}
  \hat{\omega}\sb{ij\sigma} & = &\delta\sb{ij} \left(
    \begin{array}{cc}
     (E\sb{1} + \Delta)\chi\sb{2} + a\sb{\sigma}\sp{22} & a\sb{\sigma}\sp{21}\\
     (a\sb{\sigma}\sp{21})\sp{*} & E\sb{1} \chi\sb{1} + a\sb{\sigma}\sp{22}
    \end{array} \right)
\nonumber \\
  & + & (1 - \delta\sb{ij}) V\sb{ij} \left(
    \begin{array}{cc}
      K\sb{ij\sigma}\sp{22} & K\sb{ij\sigma}\sp{21}\\
      (K\sb{ij\sigma}\sp{21})\sp{*} & K\sb{ij\sigma}\sp{11}
    \end{array} \right) ,
\label{omega}
\end{eqnarray}
while the off-diagonal matrices determine the superconducting pairing:
\begin{eqnarray}
  \hat{\Delta}\sb{ij\sigma} & = & \delta\sb{ij} \left(
    \begin{array}{cc}
      b\sb{\sigma}\sp{22} & b\sb{\sigma}\sp{21}\\
      - b\sb{\bar\sigma}\sp{21} & b\sb{\sigma}\sp{11}
    \end{array} \right)
\nonumber \\
  & + & (1 - \delta\sb{ij}) V\sb{ij} \left(
    \begin{array}{cc}
      L\sb{ij\sigma}\sp{22} & L\sb{ij\sigma}\sp{21}\\
      - L\sb{ij\bar\sigma}\sp{21} & L\sb{ij\sigma}\sp{11}
    \end{array} \right) .
\label{delta}
\end{eqnarray}
In Eq.~(\ref{omega}), the quantities $a\sp{\alpha \beta}\sb{\sigma}$ determine
energy shifts (renormalization of the chemical potential),
while the coefficients  $K\sb{ij\sigma}\sp{\alpha \beta}$ define renormalized
hopping parameters.
The site-independent anomalous correlation functions,
$\,b\sb{\sigma}\sp{\alpha \beta}\,$, and the site-dependent  ones,
$\,  L\sb{ij\sigma}\sp{\alpha \beta} \,$, in Eq.~(\ref{delta})
determine superconducting pairing in MFA.
 Explicit values of the coefficients in Eq.~(\ref{omega})
 and Eq.~(\ref{delta}) are given in the Appendix A,
 Eqs.~(\ref{sec:A7})-(\ref{sec:A18}).
\par
The QP spectrum in the normal state  of the singlet-hole Hubbard
model~(\ref{eq:H}) within the MFA   has been studied in detail by
Plakida~et~al.~\cite{Plakida95a}. Therefore here we give  only the results
of the calculations which will be used further in discussion of superconducting
pairing in the model~(\ref{eq:H}).  The QP spectrum
is described  by the normal state components of the
zero-order GF~(\ref{eq:gf0}):
\begin{equation}
  \tilde G\sb{\sigma}\sp{0}({\bf q}, \omega) =
  \left(
    \begin{array}{cc}
      \hat G\sb{\sigma}\sp{0}({\bf q}, \omega)& 0 \\
      0 & - \hat {G}\sb{\bar\sigma}\sp{0}(- {\bf q}, - \omega)\sp{(\rm T)}
    \end{array}
  \right) \, ,
\label{GFmfa}
\end{equation}
where the $\, 2\times2 \,$ matrix GF reads~\cite{Plakida95a}
\begin{equation}
  \hat {G}\sp{0}\sb{\sigma} ({\bf q}, \omega) =
  \{\omega {\hat \tau}\sb{0} - \hat {E}\sb{\sigma} ({\bf q})\}\sp{-1} \;
  \left(
    \begin{array}{cc}
      \chi\sb{2} & 0\\
      0 & \chi\sb{1}
    \end{array}
  \right) .
\label{38}
\end{equation}
The ${\bf q}$-representation of the energy matrix is given by
\begin{equation}
  \hat{E}\sb{\sigma} ({\bf q}) \ = \
  \left(
    \begin{array}{cc} \omega\sb{2}
      ({\bf q}) & W\sb{\sigma}\sp{21} ({\bf q})\\
      W\sb{\sigma}\sp{12} ({\bf q}) & \omega\sb{1} ({\bf q})
    \end{array}
  \right) \; .
\label{39}
\end{equation}
The energy spectra for the unhybridized singlet and one-hole
excitations are defined by the equations
\begin{eqnarray}
  \omega\sb{2}({\bf q}) & = & E\sb{1} + \Delta +
   a\sb{\sigma}\sp{22}/\chi\sb{2}\, +
   V\sp{22}\sb{\sigma}({\bf q})/\chi\sb{2} \, ,
\nonumber \\
  \omega\sb{1} ({\bf q}) & = & E\sb{1} + a\sb{\sigma}\sp{22}/ \chi\sb{1}\, +
   V\sp{11}\sb{\sigma} ({\bf q})/ \chi\sb{1} \, ,
\label{40}
\end{eqnarray}
while the hybridization interaction is given by
\begin{equation}
  \chi\sb{2} \, W\sp{21}\sb{\sigma} = a\sb{\sigma}\sp{21} +
   V\sp{21}\sb{\sigma} ({\bf q}), \quad
   \chi\sb{1} \, W\sp{12}\sb{\sigma} =
  \chi\sb{2} \, W\sp{21}\sb{\sigma}  .
\label{41}
\end{equation}
The effective interaction in~(\ref{40}),~(\ref{41}) has the form
\begin{equation}
  V\sp{\alpha\beta}\sb{\sigma} ({\bf q}) = \frac{t_{pd}}{N}
  \sum\sb{\bf k} \nu ({\bf k}) K\sp{\alpha\beta}\sb{\sigma} ({\bf k-q})\; ,
\label{42}
\end{equation}
where $K\sp{\alpha\beta}\sb{\sigma} ({\bf q})$ denotes the
Fourier transform of $K\sp{\alpha\beta}\sb{ij\sigma}$ in
Eq.~(\ref{omega}). Here we take into account only the nearest-
and the next-nearest neighbor hopping by writing
\begin{eqnarray}
  \nu ({\bf q}) =  2 \sum\sb{j\neq 0} \nu\sb{0j}
  {\rm e}\sp{-i{\bf q \cdot j}}
 \approx  8\nu\sb{1}\gamma ({\bf q}) +
                        8 \nu\sb{2}\gamma\sp{\prime} ({\bf q}),
\label{nu}
\end{eqnarray}
where
\begin{equation}
  \gamma ({\bf q})  =  \frac{1}{2} (\cos q\sb{x} + \cos q\sb{y}),
\quad  \gamma\sp{\prime} ({\bf q})  =  \cos q\sb{x} \cos q\sb{y} .
\label{gammann}
\end{equation}
As was shown in Ref.~\cite{Plakida95a}, the off-diagonal matrix
elements of the zero-order GF can be neglected since they
 give a small contribution of the order of $t^{12}_{ij}/ \Delta$
 to the density of states (see Fig.~2 in Ref.~\cite{Plakida95a}).
 Therefore,
  the zero-order GF~(\ref{38}) can be written in the diagonal form
\begin{equation}
  \hat G\sp{0}\sb{\sigma} ({\bf q},\omega) =
  \left(
    \begin{array}{cc}
      {\chi\sb{2}} / {[\omega - \Omega\sb{2}({\bf q})]} & 0 \\
      0 & {\chi\sb{1}} / {[\omega - \Omega\sb{1}({\bf q})]}
    \end{array}
  \right) \; ,
\label{43}
\end{equation}
where the hybridized spectra $\Omega\sb{\alpha}({\bf q})$ for
singlet $(\alpha\!=\!2)$ and for one-hole $(\alpha\!=\!1)$
excitations are given  by
\begin{eqnarray}
  \Omega\sb{2,1}({\bf q}) & = & \frac{1}{2} [\omega\sb{2} ({\bf q}) +
    \omega\sb{1} ({\bf q})] \pm \frac{1}{2} \bigl\{ [\omega\sb{2} ({\bf q}) -
    \omega\sb{1} ({\bf q})]\sp{2}
\nonumber\\
  &+ & 4 W\sp{21}\sb{\sigma} W\sp{12}\sb{\sigma} \bigr\} \sp{1/2} \, .
\label{44}
\end{eqnarray}
In these equations and in what follows,  the spin dependence of
the QP energies is omitted.
\par
To obtain a closed solution for the zero-order GF, the
correlation functions entering Eq.~(\ref{omega}) are calculated
self-consistently. The energy shifts
$a\sp{\alpha\beta}\sb{\sigma}$ depending on single-particle
correlation functions are readily calculated by using the
spectral representation of the GF~(\ref{43}). However, the
calculation of the  functions $K\sp{\alpha\beta}\sb{ij\sigma}$
depending on the density-density and spin-spin correlation
function (see Eq.~(\ref{sec:A12}))  requires the use of some
approximations.
\par
In Refs.~\cite{Beenen95,Stanescu00},  the Roth procedure was used
which involved uncontrollable  decoupling of the Hubbard
operators at the same lattice site, as e.g., $\, \langle
X\sp{02}\sb{i} X\sp{20}\sb{j} \rangle =
 \langle X\sp{0 \sigma}\sb{i} X\sp{\sigma 2}\sb{i} X\sp{20}\sb{j} \rangle \,$.
Here in the calculation of the effective hopping parameters
 $K\sp{\alpha\beta}\sb{ij\sigma}$  we apply
a simple approximation by neglecting the charge-charge
correlations $N\sb{i} N\sb{j}$ at  different lattice sites  $i
\neq j$ (as in the Hubbard I approximation) but fully take into
account  the spin-spin correlations:
\begin{equation}
  \chi\sb{ij}\sp{cs} = \frac{1}{4} \langle N\sb{i} N\sb{j} \rangle +
   \langle {\bf S}\sb{i} {\bf S}\sb{j} \rangle
   \simeq (\chi\sb{2})\sp{2} +
  \langle {\bf S}\sb{i} {\bf S}\sb{j} \rangle
\label{chics1}
\end{equation}
For a spin-singlet state without long-range magnetic order, the
GF~(\ref{43}) and the one-hole spectrum~(\ref{44}) do not depend
on spin. However, the short-range AFM spin-spin correlations are
very important and, as calculations in Ref.~\cite{Plakida95a}
have shown, the spin-spin correlation functions $\langle {\bf
S}\sb{i} {\bf S}\sb{j} \rangle$ in (\ref{chics1}) bring
significant contribution to the renormalization of the dispersion
relation~(\ref{44}). For large spin-correlations at small doping
values one finds a next-nearest neighbor dispersion. With doping,
by decreasing the spin correlations, the dispersion changes to the
 nearest neighbor one which is conventional to the overdoped case
 (see Fig.~1 in Ref.~\cite{Plakida95a}). The corresponding Fermi surface
 appears to be large even for a small doping as it is shown in
 Sec.~\ref{sec:numeric}. Therefore, the present model reproduces the most
 important features of the electronic spectra of cuprates: insulating state
 of the undoped CuO$_2$ plane and a strong reduction of the band width of
 single-particle excitations in the doped metallic state with  the
 doping dependent dispersion. Concerning a non-Fermi liquid behavior
  and a pseudogap formation in the underdoped region  one has to study
 more accurately the self-energy corrections as has been done within
 the $t$-$J$ model in Refs.~\cite{Plakida99,Prelovsek01}

\subsection{Superconducting pairing}
\label{sec:dpairing}
Let us consider the anomalous part $\hat{\Delta}\sb{ij\sigma}$,
Eq.~(\ref{delta}),  of the frequency matrix. The site independent
anomalous correlation functions $\,b\sb{\sigma}\sp{\alpha
\beta}\,$, Eqs.~(\ref{sec:A13})-(\ref{sec:A15}), vanishes  for the
$d\sb{x\sp{2}-y\sp{2}}$ pairing.  It can be easily proved by
considering the ${\bf q}$-dependent representation  of the
correlation functions, as e.g., $ \;  \sum\sb{m\neq i} V\sb{im}
\langle X\sb{i}\sp{\bar\sigma2}
   X\sb{m}\sp{\sigma 2} \rangle = \sum\sb{\bf q} V({\bf q})
  \langle  X\sp{\bar \sigma2}\sb{\bf q} X\sp{\sigma2}\sb{-{\bf q}}  \rangle
\; $. While  the interaction $V({\bf q})$ is invariant under the
permutation of the components $q\sb{x}$ and $q\sb{y}$  (in the
tetragonal phase) the anomalous correlation function  having the
$d\sb{x\sp{2}-y\sp{2}}$-wave symmetry   changes the sign that
results in vanishing of the sum over ${\bf q}$.
\par
To calculate the site dependent contributions $\,
L\sb{ij\sigma}\sp{\alpha \beta} \,$,
Eqs.~(\ref{sec:A16})-(\ref{sec:A18}),  we have to derive an
estimate of the anomalous correlation function $ \langle
X\sb{i}\sp{02} N\sb{j} \rangle $. The approach followed in
Refs.~\cite{Beenen95,Stanescu00} for its calculation employed the
Roth procedure which decouples the Hubbard operators at the same
lattice site by writing the time-dependent correlation function in
the form: $\langle c\sb{i\downarrow}(t)|c\sb{i\uparrow}(t')N\sb{j}
(t')\rangle = \langle X\sb{i}\sp{0\downarrow}(t) |
X\sb{i}\sp{\downarrow 2}(t') N\sb{j}(t') \rangle $. In view of the
identity satisfied by the Hubbard operators: $\,
X\sb{i}\sp{\alpha\beta}= X\sb{i}\sp{\alpha\gamma}
X\sb{i}\sp{\gamma\beta}\,$ for any intermediate state $|\gamma
\rangle $, the decoupling of the operators at the same lattice
site is not unique and therefore unreliable (a feature already
noticed in Refs.~\cite{Beenen95,Stanescu00}).
\par
Here we perform  a direct calculation of the correlation
function $ \langle X\sb{i}\sp{02} N\sb{j} \rangle $
{\it without any decoupling}
by considering the equation for  the corresponding commutator GF
$$
  L\sb{ij}(t-t') = \langle \langle X\sb{i}\sp{02} (t) \mid N\sb{j} (t')
                   \rangle \rangle .
$$
After differentiation with respect to the time  $t$
and use of the Fourier transform, we get the equation:
\begin{eqnarray}
  & & \left( \omega - E\sb{2} \right) L\sb{ij}(\omega) \simeq
     2 \delta\sb{ij} \langle X\sb{i}\sp{02} \rangle
\nonumber \\
  &\! +& \!\! \sum\sb{m\neq i,\sigma} \!\! 2\sigma t\sp{12}\sb{im}
     \! \left\{
        \langle\langle
          X\sb{i}\sp{0\bar\sigma} X\sb{m}\sp{0\sigma} | N\sb{j}
        \rangle\rangle\sb{\omega} \! - \!
        \langle\langle
          X\sb{i}\sp{\sigma 2} X\sb{m}\sp{\bar\sigma 2} | N\sb{j}
        \rangle\rangle\sb{\omega}
     \! \right\} \! .
\label{eq:x2a}
\end{eqnarray}
\par
In the right hand side, only the interband hopping contribution
(which mediates the exchange interaction pairing) has been
retained since the contributions from the intraband hopping
integrals give only a small renormalization  of the large
excitation energy $\, |E\sb{2}| \simeq \Delta \gg
|t\sb{ij}\sp{\alpha \alpha}|\, $. The statistical average
$\langle X\sb{i}\sp{02} N\sb{j} \rangle$ at sites $i \ne j $ can
now be evaluated from the spectral representation theorem:
\begin{eqnarray}
 & & \langle X\sb{i}\sp{02} N\sb{j} \rangle = \int\sp{+\infty}\sb{-\infty} \,
   \frac{d\omega}{1 - \exp(- \omega/ T)}\,
   \sum\sb{m\neq i,\sigma}\, 2 \sigma t\sp{12}\sb{im}
\nonumber \\
 & \times & \Bigl \{ -\frac{1}{\pi} {\rm Im}
   \Bigl [ \frac{1}{\omega - E\sb{2} + i\varepsilon}
   \Bigl ( \langle\langle X\sb{i}\sp{0\bar\sigma} X\sb{m}\sp{0\sigma} | N\sb{j}
           \rangle\rangle\sb{\omega+i\varepsilon}
\nonumber \\
 & & - \langle \langle X\sb{i}\sp{\sigma 2} X\sb{m}\sp{\bar\sigma 2} | N\sb{j}
       \rangle\rangle\sb{\omega+i\varepsilon}
   \Bigr ) \Bigr ] \Bigr \} .
\label{eq:x2b}
\end{eqnarray}
We consider below a particular case  of hole doping when the Fermi level
stays in the singlet subband: $\mu \simeq \Delta$,
and the energy parameters, $\, E\sb{2} \simeq E\sb{1} \simeq - \Delta \, $.
The contribution to the above integral coming from $\delta(\omega - E\sb{2} )$
can be neglected since it is proportional to $\, {\exp}(- \Delta/T) \ll 1$.
The contribution from the one-hole subband Green function
in Eq.~(\ref{eq:x2b}) can be estimated from its equation of motion as follows
\[
 -\frac{1}{\pi} {\rm Im}
 \langle\langle X\sb{i}\sp{0\bar\sigma} X\sb{m}\sp{0\sigma} | N\sb{j}
 \rangle\rangle\sb{\omega+i\varepsilon} \simeq \delta\sb{mj} \langle
 X\sb{i}\sp{0\bar\sigma}X\sb{j}\sp{0\sigma} \rangle \delta(\omega - 2 E\sb{1}) ,
\]
Therefore it gives also an exponentially small contribution
 $\, {\exp }(- 2\Delta/T) \ll 1 \,$.  The only non vanishing contribution  in
Eq.~(\ref{eq:x2b}) comes from the Green function of the singlet
subband:
\begin{equation}
  \langle X\sb{i}\sp{02} N\sb{j} \rangle \simeq
  - \frac{1}{\Delta }\, \sum\sb{m\neq i,\sigma} \, 2\sigma t\sp{12}\sb{im}
  \langle X\sb{i}\sp{\sigma 2} X\sb{m}\sp{\bar\sigma 2}  N\sb{j} \rangle .
\label{eq:hac}
\end{equation}
Here the  retardation effect, the frequency dependence  in the denominator
$\, 1/(\omega - E\sb{2})  \,$, is ignored since the
interband excitation energy is much larger than the QP excitation energy
in the singlet subband: $\,|E\sb{2}| \simeq \Delta \gg |t^{22}_{ij}|\,$.
The exchange interaction is usually considered in the two-site approximation,
which is obtained equating $\, m = j $ in Eq.~(\ref{eq:hac}).
In this approximation, we get after summation over the spin $\sigma$:
\begin{equation}
  \langle X\sb{i}\sp{02} N\sb{j} \rangle
  = \langle c\sb{i\downarrow}c\sb{i\uparrow} N\sb{j}\rangle
 = - \frac{4t\sp{12}\sb{ij} }{\Delta}
   2 \sigma \, \langle X\sb{i}\sp{\sigma2}X\sb{j}\sp{\bar\sigma2}\rangle .
 \label{eq:x2d}
\end{equation}
where we have used the identity for the Hubbard operators, $\,
X\sb{j}\sp{\bar\sigma 2}  N\sb{j} = 2 X\sb{j}\sp{\bar\sigma 2},
\,$ and Eq.~(\ref{eq:x2da}).
\par
 This finally allows us to write the expressions
of the anomalous component $\Delta^{22}_{ij\sigma}  $
for the case of hole doping as follows
\begin{eqnarray}
\Delta^{22}_{ij\sigma} & = &  \frac{1}{N}\sum_{\bf q} {\rm e}^{i{\bf
q\cdot (i - j)}} \; \Delta^{22}_{\sigma}({\bf q}) = J_{ij}  \langle
X_{i}^{\sigma2}X_{j}^{\bar\sigma2}\rangle \, .
 \label{eq:x2e}
\end{eqnarray}
This result recovers the exchange interaction contribution to the pairing,
with an exchange energy parameter
$\, J_{ij} = {4\, (t^{12}_{ij})^2}/{\Delta}$.
The anomalous component $\, \Delta^{11}_{ij\sigma}  \,$
for the case of hole doping can be neglected since its contribution
to the gap equation is extremely small, of the order of
$\Delta^{11}_{\sigma}({\bf q})/ \Delta $.
\par
In the case of electron doping, on the contrary, we can neglect the
anomalous correlation function for the singlet subband,
$\, \Delta^{22}_{ij\sigma}$, while  an analogous calculation
for the anomalous correlation function of the one-hole subband gives
\begin{eqnarray}
\Delta_{ij\sigma}^{11}&=&
 J_{ij}\, \langle X_{i}^{0\bar\sigma} X_{j}^{0\sigma} \rangle .
\label{eq:x2el}
\end{eqnarray}
\par
 We have therefore proved that in the
MFA the interband  anomalous correlation function is of the order
of $t\sp{12}\sb{ij}/ \Delta $ and it is proportional to the
statistical average of the conventional electron (hole) pairs at
neighboring lattice sites in the  Hubbard subband which intersects
the Fermi level. As a consequence,  the anomalous contributions to
the zero-order GF, Eq.~(\ref{eq:gf0}), originate in conventional
anomalous pairs of quasi-particles and  their  pairing in MFA is
mediated by the exchange interaction which has been  studied in
the $t$-$J$ model (see, e.g.,~\cite{Plakida89,Plakida99}). In view
of this conclusion, the MFA nonzero superconducting pairing
reported in the frame of the conventional Hubbard
model~\cite{Beenen95,Avella97,Stanescu00} can be inferred to stem
from the exchange interaction which equals to $\, J\sb{ij} = {4
t\sp{2}}/{U}$ in this model. The exchange interaction vanishes in
the limit $\, U \to \infty\,$, a feature which explains the
disappearance of the pairing at large
$U$.~\cite{Beenen95,Avella97,Stanescu00}
\par
To summarize the study of the MFA, let us write down a closed system of
equations  for the particular case of
hole doping with the chemical potential in the singlet subband. The
$2\times 2$ matrix GF for the singlet subband writes:
\begin{equation}
  \hat G\sp{22}\sb{ij, \sigma}(\omega) =
  \langle\langle
    { X\sb{i}\sp{\sigma 2} \choose X\sb{i}\sp{2 \bar\sigma} } \mid
    ( X\sb{j}\sp{2 \sigma}\,X\sb{j}\sp{\bar{\sigma} 2})
  \rangle\rangle\sb{\omega}\, .
\label{G22}
\end{equation}
By taking into account the normal part of the GF in the diagonal form,
Eq.~(\ref{43}), and the anomalous part of the frequency matrix for the
singlet subband,  Eq.~(\ref{eq:x2e}), we can write the GF~(\ref{G22}) in
the form:
\begin{equation}
  \hat G\sp{22(0)}\sb{\sigma} ({\bf q},\omega) = \chi\sb{2}
  \left\{
    \omega \hat\tau\sb{0} - \Omega\sb{2}({\bf q})\hat \tau\sb{3} -
    \phi\sb{\sigma}\sp{22}({\bf q}) \hat \tau\sb{1}
  \right\}\sp{-1} ,
\label{G22a}
\end{equation}
where $\hat \tau_{0}, \; \hat \tau_{1}, \; \hat \tau_{3} $ are the Pauli
matrices. Here we introduced the gap function in MFA for the singlet subband
$\,\phi_{\sigma}^{22}({\bf q})=\Delta^{22}_{\sigma}({\bf q}) /\chi_{2} \,$
induced by the exchange interaction. From  Eq.~(\ref{eq:x2e}) and
equation for the GF~(\ref{G22a}) we get a self-consistent  BCS-type
equation for the gap function in the MFA:
\begin{equation}
  \phi\sb{\sigma}\sp{22}({\bf q}) = \frac{1}{N} \sum\sb{\bf k} J({\bf k-q})
  \frac{\phi\sb{\sigma}\sp{22}({\bf k})}{2 {\cal E}\sb{2}({\bf k})}
  \tanh\frac{{\cal E}\sb{2}({\bf k})}{2T},
\label{phi}
\end{equation}
where $J({\bf q}) = 4 J\gamma({\bf q})$ is the Fourier component
of the nearest neighbor exchange interaction and $\, {\cal
E}\sb{2}({\bf k}) = [\Omega\sb{2}({\bf k})\sp{2} +
|\phi\sb{\sigma}\sp{22}({\bf k})|\sp{2}]\sp{1/2} \,$ denotes the
QP energy.
  Analogous equations can be obtained
for the electron doped case, $n < 1$, with the chemical potential in the
one-hole band by considering GF for the Hubbard operators $\,
X\sb{i}\sp{0\sigma}, \; X\sb{i}\sp{\bar\sigma 0}\,$.
\par
The  obtained results for the gap equation in MFA for
superconducting pairing mediated by AFM exchange interaction  are
identical to the MFA results for the $t$-$J$ model (see,
e.g.,~\cite{Plakida89,Plakida99}). However, we have derived these
results from the original two subband Hubbard model~(\ref{eq:H})
which has allowed an understanding of the role of retardation
effects in the exchange pairing. We have proved that  they can be
neglected due to the large interband excitation energy in
comparison with the QP intraband excitation energy. We also
obtained a more general, three-site representation for the
anomalous correlation function~(\ref{eq:hac}) which can be used to
study the many-particle exchange pairing.

\section{Self-energy corrections}
\label{sec:s-e}
\subsection{Self-consistent Born approximation}
\label{sec:scba}

To study the finite life-time effects and renormalization of the
QP spectra caused by inelastic scattering  one should calculate
the self-energy~(\ref{eq:self-enir-qo}) induced by many-particle
excitations. A general representation for the self-energy in the
$({\bf r}, \omega)$-representation  reads
\begin{equation}
  \tilde \Sigma\sb{ij\sigma}(\omega) = \tilde\chi\sp{-1}
   \left(
    \begin{array}{cc}
      \hat M\sb{ij\sigma}(\omega)& \hat\Phi\sb{ij\sigma}(\omega)\\
      \hat\Phi\sb{ij\sigma}\sp{\dagger ({\rm T})}(\omega) &
      - \hat{M}\sb{ij\bar\sigma}\sp{({\rm T})}(-\omega)
    \end{array}
   \right) \tilde \chi\sp{-1} \, ,
\label{eq:sigma-ro}
\end{equation}
where the $2\times 2$ matrices $\hat M$ and $\hat\Phi$ denote the
normal and anomalous contributions to the self-energy,
respectively (see Appendix B,
Eqs.~(\ref{sec:B4}),~(\ref{sec:B5})).
\par
 Here we consider the self-consistent Born
approximation (SCBA) (or the non-crossing approximation) which is
proved to be quite reliable in studies of the $t$-$J$
model.~\cite{Plakida99} In SCBA, the propagation of the Fermi-like
excitation, $X\sb{1}(t),\,$ and Bose-like excitation, $
B\sb{1'}(t),\,$ (spin or charge - see Appendix A,
Eq.~(\ref{sec:A4}))  in the many-particle GF
in~(\ref{eq:sigma-ro}) are assumed to be uncorrelated. In the
diagram technique SCBA is represented by a skeleton loop diagram
without vertex corrections. Therefore the SCBA results from the
decoupling of the corresponding operators in the many-particle
time-dependent correlation functions for lattice sites $\,( 1 \neq
1', \; 2 \neq 2')\,$ as follows:
\begin{eqnarray}
  & &\langle B\sb{1'}(t) X\sb{1}(t) B\sb{2'}(t') X\sb{2}(t') \rangle
\nonumber \\
  & \simeq & \langle X\sb{1}(t) X\sb{2}(t') \rangle
   \langle B\sb{1'}(t) B\sb{2'}(t') \rangle.
\label{eq:decoupl-ro1}
\end{eqnarray}
Using the spectral theorem, the SCBA results in the following
decoupling relation for the many-particle GF:
$$
  \langle\!\langle B\sb{1'} X\sb{1} \!\mid\! B\sb{2'} X\sb{2}
  \rangle\!\rangle\sb{\omega} \simeq \frac{1}{  \pi\sp{2}}\!
  \int\limits\sb{-\infty}\sp{+\infty} \int\limits\sb{-\infty}\sp{+\infty}\!\!
  \frac{{\rm d}\omega\sb{1} {\rm d}\omega\sb{2}}
       {\omega - \omega\sb{1} - \omega\sb{2}}
$$
\begin{equation}
  \times \, N(\omega\sb{1},\omega\sb{2})\;
  \mbox{Im} \langle\!\langle X\sb{1} \!\mid\!
         X\sb{2} \rangle\!\rangle\sb{\omega\sb{1}}
  \mbox{Im}\langle\!\langle B\sb{1'}\!\mid\!
         B\sb{2'}\rangle\!\rangle\sb{\omega\sb{2}} ,
\label{eq:decoupl-ro}
\end{equation}
where $\, N(\omega\sb{1},\omega\sb{2}) =(1/2)
 [\tanh (\omega\sb{1}/ 2T ) + \coth (\omega\sb{2}/2T)].\,$
 Within this approximation we obtain a self-consistent system
of equations for the self-energy~(\ref{eq:sigma-ro}) and
single-particle GF~(\ref{eq:gf2x2}). However, to get a tractable
problem, we simplify this system by using a diagonal
approximation for the GF as has been assumed for the zero-order
GF~(\ref{G22a}). Consequently, keeping  only the diagonal
self-energy components, $\,M\sb{\sigma}\sp{\alpha\alpha}\,$ and
$\,\Phi\sb{\sigma}\sp{\alpha\alpha}\,$ in Eq.~(\ref{eq:sigma-ro}),
 we can solve the $4\times 4$
matrix Dyson equation~(\ref{eq:Dyson}) in the form of two
independent $2\times 2$ matrices  for GF for the singlet and
one-hole subbands. The diagonal approximation gives the lowest
order contribution to the self-energy in the small hybridization
parameter $\,(t\sp{12}\sb{ij}/\Delta)$.
\par
By taking into account the zero-order $2\times 2$ matrix  GF for
the singlet subband, Eq.(\ref{G22a}),  we derive the following
Dyson equation for   GF~(\ref{G22}):
\begin{eqnarray}
  \hat G\sp{22}\sb{\sigma} ({\bf q},\omega) & = & \chi\sb{2}
  \{   \omega \hat\tau\sb{0} - [\Omega\sb{2}({\bf q})+
    M\sb{\sigma}\sp{22}({\bf q},\omega)]\hat \tau\sb{3}
\nonumber\\
  & - & [\phi\sb{\sigma}\sp{22}({\bf q})+
    \Phi\sb{\sigma}\sp{22}({\bf q},\omega)] \hat \tau\sb{1}\}\sp{-1} .
    \label{G22b}
\end{eqnarray}
Here the normal, $\,  M\sb{\sigma}\sp{22}({\bf q},\omega) \,$, and
anomalous, $\,  \Phi\sb{\sigma}\sp{22}({\bf q},\omega) \,$, parts
of the self-energy for the singlet subband in the SCBA
approximation can be written in the form:
\begin{eqnarray}
   M\sb{\sigma}\sp{22}({\bf q},\omega) = \frac{1}{N} \sum\sb{\bf k}
   \int\limits\sb{-\infty}\sp{+\infty} \!\!{\rm d}\omega\sb{1}
    K\sp{(+)}(\omega,\omega\sb{1}|{\bf k},{\bf q - k})
\nonumber \\
  \times \; \left\{ - \frac{1}{\pi} \mbox{Im} \left[  K\sb{22}\sp{2}
   G\sb{\sigma}\sp{22}({\bf k},\omega\sb{1}) +  K\sb{12}\sp{2}
   G\sb{\sigma}\sp{11}({\bf k},\omega\sb{1}) \right] \right\} \; ,
\label{eq:m-qo} \\
  \Phi\sb{\sigma}\sp{22}({\bf q},\omega) = \frac{1}{N} \sum\sb{{\bf k}}
   \int\limits\sb{-\infty}\sp{+\infty} \!\!{\rm d}\omega\sb{1}
    K\sp{(-)}(\omega,\omega\sb{1}|{\bf k},{\bf q - k})
\nonumber \\
  \times \; \left\{ - \frac{1}{\pi} \mbox{Im} \left[  K\sb{22}\sp{2}
   F\sb{\sigma}\sp{22}({\bf k},\omega\sb{1}) -  K\sb{12}\sp{2}
   F\sb{\sigma }\sp{11}({\bf k},\omega\sb{1}) \right] \right\} .
 \label{eq:phi-qo}
\end{eqnarray}
The kernel of the integral equations for the self-energy is
defined by the equation
\begin{eqnarray}
  K\sp{(\pm)}(\omega,\omega\sb{1}|{\bf k},{\bf q - k}) =
  t\sp{2}\sb{pd}\, {|\nu({\bf k})|\sp{2}}
\nonumber \\
\times \, \int\limits\sb{-\infty}\sp{+\infty}\;
   {\rm d}\omega\sb{2}\; \frac{ N(\omega\sb{1},\omega\sb{2})}
   {\omega - \omega\sb{1} - \omega\sb{2}}
 \left[ \,  \frac{1}{\pi} \,  \mbox{Im} \,
 \chi\sp{(\pm)}\sb{sc}({\bf q -k},\omega\sb{2})
  \right]  ,
\label{eq:Kpm}
\end{eqnarray}
where $\nu ({\bf k})$ is given by Eq.~(\ref{nu}). The spectral
density of spin-charge fluctuations is defined by the
corresponding dynamical susceptibilities which are given by the
commutator GF
\begin{eqnarray}
 \chi\sp{(\pm)}\sb{sc}{\bf q},\omega)
   & = & \chi\sb{s}({\bf q},\omega) \pm \chi\sb{c}({\bf q},\omega)
 \nonumber \\
 & = &
   \langle\!\langle {\bf S\sb{q} | S\sb{-q}} \rangle\!\rangle\sb{\omega}
  \pm \frac{1}{4} \langle\!\langle \delta N\sb{\bf q} | \delta N\sb{-\bf q}
   \rangle\!\rangle\sb{\omega} \, .
\label{eq:D-qo}
\end{eqnarray}
 Analogous expression for the GF in terms of the Hubbard operators
  $\, X\sb{i}\sp{0\sigma}, \; X\sb{i}\sp{\bar\sigma 0}\,$,  and the
corresponding self-energy components can be written for the
one-hole subband in the case of electron doping (see Appendix~B,
Eqs.~(\ref{sec:B6})-(\ref{sec:B9})).

\subsection{Weak coupling approximation}
\label{sec:w-c}

Self-consistent numerical solution of the coupled system of
equations for the GF~(\ref{G22b}) and the
self-energy~(\ref{eq:m-qo}),~(\ref{eq:phi-qo}) is  rather
complicated (see, e.g. calculations in Ref.~\cite{Plakida99}). To
estimate the role of the kinematic interaction in superconducting
pairing and its contribution to superconducting $T_c$ beyond the
MFA,   we study below a simplified approach  based on the weak
coupling approximation (WCA). In  WCA, it is assumed that the
behavior of an electron liquid  is dominated by the interactions
around the Fermi level and therefore the interaction
kernel~(\ref{eq:Kpm}) at frequencies $\,(\omega, \omega\sb{1})\,$
close to the Fermi surface (FS) can be  factorized in the form
\begin{eqnarray}
  & & K\sp{(\pm)}(\omega,\omega\sb{1}|{\bf k},{\bf q - k})
\nonumber\\
  & \simeq & - \frac{1}{2} \tanh \left( \frac{\omega\sb{1}}{2 T} \right)
      \lambda\sp{(\pm)}({\bf k}, {\bf q - k}) \, ,
\label{eq:Kpm-wc}
\end{eqnarray}
for $\, |\omega, \omega\sb{1}| \le \omega\sb{s} \ll W \,$ where
$\, \omega\sb{s} $
is a characteristic pairing energy and $\, W \,$ is the band width.
In this approximation, the effective interaction is defined by the
static susceptibility
\begin{eqnarray}
  &&\lambda\sp{(\pm)}({\bf k},{\bf q - k}) = t\sp{2}\sb{pd}\;
  |\nu({\bf k})|\sp{2}
\nonumber\\
  &&\times \; \int\limits\sb{-\infty}\sp{+\infty}
    \frac{{\rm d}\omega\sb{2}} {\omega\sb{2}}
    [ \, \frac{1}{\pi}\, \mbox{Im} \,
    \chi\sp{(\pm)}\sb{sc}({\bf q - k},\omega\sb{2})]
\nonumber\\
  & = &  t\sp{2}\sb{pd} \; |\nu({\bf k})|\sp{2}
   \,  \mbox{Re} \, \chi\sp{(\pm)}\sb{sc} ({\bf q - k},\omega\sb{2} =0)] \; .
\label{eq:lambda}
\end{eqnarray}
The WCA is suitable for the band which crosses the FS.  For the
another band, which is far away from the FS at an energy of the
order of the band gap, $\, \omega\sb{1} \simeq \Delta$, an
integration over $\omega\sb{1}$ in Eqs.(\ref{eq:m-qo})
and~(\ref{eq:phi-qo}) is straightforward.
  \par
For a hole doped  system, $n= 1+\delta\geq 1 $, the
chemical potential is in the singlet band, $\mu \simeq \Delta$,
and we can write the dispersion relations for the two bands in
the normal state as follows
\begin{eqnarray}
  \bar\Omega\sb{2}({\bf q})&=&\Omega\sb{2}({\bf q}) +
    \chi\sb{2}\sp{-1} M\sb{\sigma}\sp{22}({\bf q},
      \omega=\bar\Omega\sb{2}({\bf q}))
\nonumber\\
& \simeq & \Delta -\mu + \epsilon\sb{2}({\bf q})
  \simeq \epsilon\sb{2}({\bf q}) \; ,
\label{g9a}
\end{eqnarray}
\begin{eqnarray}
\bar\Omega\sb{1}({\bf q})&=&\Omega\sb{1}({\bf q}) +
    \chi\sb{1}\sp{-1} M\sb{\sigma}\sp{11}({\bf q},
      \omega=\bar\Omega\sb{1}({\bf q}))
\nonumber\\
  & \simeq & -\mu + \epsilon\sb{1}({\bf q})
  \simeq -\Delta + \epsilon\sb{1}({\bf q})\, ,
\label{g9b}
\end{eqnarray}
where the MFA energy $\, \Omega\sb{\alpha}({\bf q}) \,$ is
defined by Eq.~(\ref{44}).  The zero of the energy in the singlet
band is fixed at the Fermi wave-vector $\; \epsilon\sb{2}({\bf
q}\sb{F})= 0 \;$.
\par
Integration over $\omega\sb{1}$ in~(\ref{eq:phi-qo}) gives the
following  equation for the  anomalous component of the
self-energy
\begin{eqnarray}
  \Phi\sp{22}\sb{\sigma}({\bf q},\omega  \simeq 0) =  - K\sb{22}\sp{2}
   {\cal S}\sb{2,\sigma}({\bf q})
    + K\sb{12}\sp{2} {\cal S}\sb{1,\sigma}({\bf q}) .
\label{eq:phiweak}
\end{eqnarray}
The sum ${\cal S}\sb{2,\sigma}({\bf q})$ for the singlet band at
the FS is given by
\begin{equation}
 \! {\cal S}\sb{2,\sigma}({\bf q}) = \frac{1}{N} \sum\sb{{\bf k}}
   \lambda\sp{(-)}({\bf k}, {\bf q - k})
 \frac{\bar\Phi\sp{22}\sb{\sigma}({\bf k})}
  {2 {\bar{\cal E}}\sb{2}(\bf k)}
   \tanh \frac{{\bar{\cal E}}\sb{2}(\bf k)}{2 T},\!\!
\label{eq:sumweak}
\end{equation}
where $\, \bar\Phi\sp{22}\sb{\sigma}({\bf k})=
\phi\sp{22}\sb{\sigma}({\bf k}) +
   \Phi\sp{22}\sb{\sigma}({\bf q},0)\,\chi\sb{2}\sp{-1}\,$ is
an effective gap function.  The integration over ${\bf k}$ is
restricted here to an energy shell around the Fermi energy of the
order of a characteristic energy $\, \omega\sb{s} \,$ of the spin
(charge) fluctuations defined by the
susceptibility~(\ref{eq:D-qo}). The quasiparticle energy is given
by
\begin{equation}
 {\bar{\cal E}}\sb{2}({\bf q}) = \left[ \bar \Omega\sb{2}\sp{2}({\bf q}) +
  \mid \bar\Phi\sp{22}\sb{\sigma}({\bf q}) \mid \sp{2} \right]
  \sp{1/2}.
\label{eq:e12}
\end{equation}
The one-hole band lies below the FS at an energy of the order
$\Delta \gg W$.
Therefore, integration over $\omega\sb{1}$ in Eq.~(\ref{eq:phi-qo})
for the anomalous GF
$\,  F\sb{\sigma }\sp{11}({\bf k},\omega\sb{1})\,$ can  be done
by neglecting for  the QP excitation energy  in this band
the dispersion in  Eq.~(\ref{g9b}) as well as the superconducting gap:
 $\quad    F\sb{\sigma }\sp{11}({\bf k},\omega\sb{1})\;
  \simeq \;\phi\sp{11}\sb{\sigma}({\bf k})\;/ \,
  (\omega\sb{1}\sp{2} - \Delta\sp{2} )  \,$.\\
 This results in the estimate
\begin{equation}
  {\cal S}\sb{1,\sigma}({\bf q}) = \frac{1}{N} \sum\sb{\bf k}
  t\sp{2}\sb{pd} |\nu({\bf k})|\sp{2} \frac{\phi\sp{11}\sb{\sigma}({\bf k})}
  {\Delta \sp{2}}\, L\sb{sc}\sp{(-)}({\bf k - q}).
   \label{eq:sumweak1}
\end{equation}
In this equation, the spin-charge static correlation function
$\;
  L\sb{sc}\sp{(-)}({\bf q}) = \langle {\bf S\sb{q}} {\bf S\sb{-q}}\rangle
  - \frac{1}{4} \langle \delta N\sb{\bf q} \delta N\sb{-\bf q} \rangle \; $
resulted from the integration over $\omega\sb{2}$ in
Eq.~(\ref{eq:Kpm}). A simple estimate shows that the sum $\,
{\cal S}\sb{1,\sigma}({\bf q}) \,$ gives a small contribution, of
the order $\, (t\sb{eff}/ \Delta)\sp{2} \,
\phi\sp{11}\sb{\sigma}({\bf k}) \simeq 10\sp{-2}
\,\phi\sp{11}\sb{\sigma}({\bf k})\, $, and it can be neglected in
Eq.~(\ref{eq:phiweak}).
\par
By taking into account the contribution due to the exchange
interaction in MFA, Eq.~(\ref{phi}), the equation for the
effective  gap in the singlet subband  can be written as follows:
\begin{eqnarray}
 \bar\Phi\sp{22}\sb{\sigma}({\bf q}) & = & \frac{1}{N} \sum\sb{\bf k} \left[ J({\bf k - q})
     - K\sb{22}\sp{2}\, \lambda\sp{(-)}({\bf k}, {\bf q - k}) \right]
\nonumber \\
  && \times \frac{\bar\Phi\sp{22}\sb{\sigma}({\bf k})}
  {2{\bar{\cal E}}\sb{2}(\bf k)}
     \tanh \frac{{\bar{\cal E}}\sb{2}(\bf k)}{2 T} \; .
\label{gap22}
\end{eqnarray}
In this equation integration over $\bf k$ for the exchange interaction is
performed without any restriction on the QP energy,
while for the spin-fluctuation contribution,
$\, \lambda\sp{(-)}({\bf k}, {\bf q - k}) ,\,$ the energy of the
pairing quasi-particles  is confined to a narrow energy shell
of the order of $\omega_s$ close to the FS as pointed out in
Eq.~(\ref{eq:Kpm-wc}).
\par
Similar considerations hold true for an electron doped system,
$\,n= 1+\delta\leq 1\,$ when $\,\mu \simeq 0$ in~(\ref{g9b}).  In
that case, the significant WCA equation involves the gap $\,
\bar\Phi\sp{11}\sb{\sigma}({\bf q}) \,$ and  critical $T\sb{c}$
defined by the corresponding equation analogous to
Eq.~(\ref{gap22}).
\par
 Thus,     we have proved that in the
strong correlation limit of the effective Hubbard
model~(\ref{eq:H}), the equations of the gap functions of the  two
subbands: singlet and one-hole,  can be solved independently
since the self-energy corrections (both the normal and anomalous
components) from the another subband give a vanishingly small
contribution due to the large interband gap $\Delta \gg W $.

\section{Numerical results and discussion}
\label{sec:numeric}
We start with  an analytical estimation of the superconducting
$T\sb{c}$ mediated by the exchange interaction and spin
fluctuations in the gap equation~(\ref{gap22}) for a hole doped
case.  The charge-charge fluctuations contribution is small due
to their large energy as compared to the spin-spin fluctuations
and they are neglected. Therefore the effective interaction
mediated by spin fluctuations in Eq.~(\ref{eq:lambda}) can be
written in the form
\begin{equation}
  \lambda\sb{s}({\bf k}, {\bf q - k}) = t\sb{pd}\sp{2} \,|\nu({\bf k})|\sp{2}
\, \chi\sb{s}({\bf q-k}) \, ,
 \label{eq:lambda1}
\end{equation}
where $\, \chi\sb{s}({\bf q-k})\,$ is the static spin
susceptibility.  It can be evaluated from  a dynamical spin
susceptibility suggested in numerical studies~\cite{Jaklic95} as
follows:
\begin{equation}
 \chi\sb{s}({\bf q}) \simeq  \frac{1}{\omega_s}
   \langle {\bf S\sb{q}} {\bf S\sb{-q}} \rangle
=  \frac{\chi_0(\xi)}{1 + \xi\sp{2} [1 + \gamma({\bf q})] } .
\label{eq:oimchi}
\end{equation}
There are two fitting parameters in the model: a short-range AFM
correlation length $\xi$  of the order of a few lattice spacing
and a cut-off energy of spin-fluctuations $\omega\sb{s} \,$ of the
order of the exchange energy $\, J$.  The spin
susceptibility~(\ref{eq:oimchi})  has a maximum value $\chi_0
(\xi)$ at the AFM wave vector ${\bf Q}=(\pi,\pi)$ when
$1+\gamma({\bf Q})= 0 .\, $ Its value is fixed by the
normalization condition for $n = 1+ \delta \geq 1$:
\begin{equation}
\frac{1}{N}\sum_{i} \langle {\bf S_i}{\bf S_i}\rangle =
 \frac{1}{N} \sum\sb{\bf q}
  \langle {\bf S}\sb{\bf q}{\bf S}\sb{\bf -q} \rangle  =
  \frac{3}{4} (1-\delta) ,
\label{eq:norm}
\end{equation}
which gives $\;\chi_0(\xi) = 3(1-\delta)/(4\, \omega_s\, C(\xi))
\;$  where $\; C(\xi) = (1/N)\, \sum\sb{\bf q} \,
  \{ 1 + \xi\sp{2} [1 + \gamma({\bf q})]\}\sp{-1} .\;$
At large $\xi$, $\, \chi_0(\xi)  \propto \xi\sp{2} / \ln\xi \,$.
\par
For an analytical estimation of superconducting $T\sb{c}$ we
consider the linearized  Eq.~(\ref{gap22})  where for the gap we
assume the $d\sb{x\sp{2}-y\sp{2}}$-wave  symmetry in the
conventional  form:
\begin{equation}
 \bar{\Phi}\sp{22}\sb{\sigma}({\bf q}) =
  \phi\sp{22}\sb{\sigma}(\cos q\sb{x} - \cos q\sb{y})=
  \phi\sp{22}\sb{\sigma} \eta({\bf q}) .
\label{g1}
\end{equation}
Then integrating over ${\bf q}$  both sides of Eq.~(\ref{gap22})
multiplied by $\, \eta({\bf q})\, $ we get the following equation
for $\, T\sb{c}$:
\begin{eqnarray}
 1 & = & \frac{1}{N} \sum\sb{\bf k}
 \frac{1} {2 \bar \Omega\sb{2}({\bf k})}
  \tanh \frac{\bar \Omega\sb{2}({\bf k})}{2 T\sb{c}}
\nonumber \\
&& \times \left[ J\,\eta({\bf k})\sp{2}  + \lambda\sb{s}
      \,(4\gamma({\bf k}))\sp{2}
       \eta({\bf k})\sp{2} \right]  \; .
\label{gap2}
\end{eqnarray}
For the exchange interaction $\,J({\bf q - k})= 4J \gamma({\bf q
- k})\,$ integration over ${\bf q}$ gives $\,(4J/N) \sum\sb{\bf
q} \, \gamma({\bf k - q})\, \eta({\bf q})\, = J \eta({\bf k})
\sp{2}\, $. In ${\bf q}$-integration of the static spin
susceptibility in Eq.~(\ref{eq:lambda1}) we have taken into account
that it is positive and shows a strong peak  at the AFM wave-vector
${\bf q - k = Q} = (\pi,\pi)$. Therefore  the sum over ${\bf q}$
 was evaluated at this wave vector: $\, (1/N)
\sum\sb{\bf q} \eta({\bf q}) \chi\sb{s}({\bf q - k}) \simeq
-\eta({\bf k}) \sum\sb{\bf q'}  \chi\sb{s}({\bf q'}) = -\eta({\bf
k})\, 3(1-\delta)/4\omega_s \,$. The effective coupling constant
for spin-fluctuation pairing in Eq.(\ref{gap2}) is  given by $\,
\lambda\sb{s} = (K\sb{22}\, t\sb{pd} \,2\nu_1)\sp{2}3 (1-\delta)/
4\omega_s \simeq t\sp{2}\sb{eff}/ \omega_s\,$.
\par
 Now we can perform integration over the QP energy $\,
\Omega\sb{2}({\bf k})\,$ in Eq.~(\ref{gap2}) by introducing an
effective density of electronic states (DOS) for two interactions:
\begin{eqnarray}
N\sb{d}(\varepsilon) & = & \frac{1}{N}\sum\sb{\bf k}\,
  \eta({\bf k})\sp{2}\delta(\varepsilon - \bar \Omega\sb{2}({\bf
  k}))
  \,, \label{dos1}\\
N\sb{sf}(\varepsilon) & = & \frac{1}{N}\sum\sb{\bf k}\,
  \eta({\bf k})\sp{2}\, (4\gamma({\bf k}))\sp{2}\,
  \delta(\varepsilon - \bar \Omega\sb{2}({\bf k}))\, .
\label{dos2}
\end{eqnarray}
They are normalized to unity since $\,({1}/{N})\sum\sb{\bf k}\,
  \eta({\bf k})\sp{2}= 1\,$ and $\,({1}/{N})\sum\sb{\bf k}\,
  \eta({\bf k})\sp{2}\,(4\gamma({\bf k}))\sp{2} = 1\,$.
 The  resulting  equation for $T\sb{c}$ takes the form:
\begin{eqnarray}
1 & \simeq &\int\sb{-\mu}^{\widetilde{W} -\mu} \frac{d
\epsilon}{2\epsilon} \tanh \frac{\epsilon}{2T_c}
\nonumber\\
&\times& [J N_{d}(\epsilon)+ \theta(\omega_s -
|\varepsilon|)\lambda\sb{s} N\sb{sf}(\varepsilon)] ,
\label{gap3}
\end{eqnarray}
where $\,\theta(\omega_s - |\varepsilon|) =1\, $ for QP energy $\,
|\varepsilon| < \omega_s \,$ and equals zero otherwise. Here $\,
\widetilde{W}  \,$ is a renormalized band width for the singlet
subband, Eq.~(\ref{g9a}).
\par
Let us  compare  superconducting pairing mediated by the exchange
interaction and the spin-fluctuations. While for the exchange
interaction - the first term in Eq.~(\ref{gap3}), the integration
over energy is performed over the whole conduction band, in the
spin-fluctuation contribution - the second term in
Eq.~(\ref{gap3}), the integration over energy is restricted to a
narrow energy shell close to the FS: $\,\omega_s \ll E_F .\,$
Moreover,  the effective DOS~(\ref{dos1}) for the exchange
interaction includes  large contribution from states near the van
Hove singularity at $(\pi,0)$ points, while for the
spin-fluctuation interaction the corresponding DOS~(\ref{dos2}) is
strongly suppressed since  the kinematic interaction which is
proportional to $\, \nu({\bf k})\simeq 8\nu_1 \gamma({\bf k})\, $
vanishes along the lines $\,|k_x|+|k_y| = \pi \,$. The effect of
vanishing of spin-fluctuation pairing close to the AFM Brillouin
zone, $\,|k_x|+|k_y| = \pi ,\,$ was first mentioned by
Schrieffer~\cite{Schrieffer95} in his remarks concerning the model
of nearly AFM Fermi liquid.~\cite{Monthoux93}
\par
These specific
properties of interactions result in a lower $ T_c $ mediated by
spin-fluctuations in comparison with the exchange interaction
though the coupling constant are comparable: $\, J \simeq
0.15$~eV, while
 $\, \lambda\sb{s} \simeq t\sp{2}\sb{eff}/ \omega_s \simeq
 0.27$~eV for $\, t\sb{eff}\simeq 0.2$~eV and $\,\omega_s
 \simeq 0.15$~eV. To demonstrate this we solve
Eq.~(\ref{gap3})  by applying the conventional logarithmic
approximation  since  $\, T\sb{c} \ll \omega_s \ll   \mu \,$.
Then we get the following estimate for $T\sb{c}$ mediated by  the
exchange interaction
\begin{equation}
 T\sb{c}\sp{ex} \simeq 1.14 \, \sqrt{\mu ( \widetilde{W}-\mu)}
 {\exp}{(-1/J \, N_d(\delta))} ,
 \label{tcJ}
\end{equation}
where  $\, N_{d}(\delta) \,$  is an average DOS~(\ref{dos1}) for a
hole concentration $ \delta $. The superconducting
$T\sb{c}\sp{ex}\, $ peaks at an optimal doping for the chemical
potential $\,  \mu = \widetilde{W}/2 \,$ and being proportional to
a large electronic energy $\,\mu = E_F\, $ even for a weak
coupling reaches  a high value. For instance, by assuming the
model parameters: $\, E_F = \widetilde{W}/2 \simeq 0.35$~eV, $\, J
\simeq 0.15$~eV, and $N_{d}(\delta)\simeq 2\; ({\rm eV \cdot
spin})\sp{-1}$, we get $\,  J \, N(\delta) \simeq 0.3 \,$ and $\,
 T_c \simeq 170$~K.
 \par
 For the spin-fluctuation pairing we obtain the BCS-like formula
 \begin{equation}
T\sb{c}\sp{sf} \simeq 1.14 \,\omega_s \,
 {\exp}{(-1/ \lambda\sb{s} \, N_{sf}(0)}) ,
 \label{tcsf}
\end{equation}
where $\, N_{sf}(0) \,$ is DOS~(\ref{dos2}) at the FS. For a large
FS, close to the lines $\,|k_x|+|k_y| = \pi \,$, it is rather
small as pointed out above that results in a weak effective
coupling. Therefore $T\sb{c}\sp{sf}$, being proportional to
$\omega_s \ll \mu $ and having weaker coupling should be lower
than $T\sb{c}\sp{ex}$, Eq.~(\ref{tcJ}). It should be stressed that
in comparison with phenomenological models for spin-fluctuation
pairing, as e.g. in Ref.~\cite{Monthoux93}, the coupling constant
$\, \lambda\sb{s} \, N_{sf}(0) \, $ in Eq.(\ref{tcsf}) is  defined
by the original microscopic parameters of the Hubbard
model~(\ref{eq:H}).
\par
Similar estimations can be done for the electron doped case, $n=
1+\delta\leq 1$, where $\mu \simeq 0$. In this case, the
dominating contribution to $T\sb{c}$ comes from the integral
${\cal S}\sb{1}$. For a conventional  Hubbard model with only one
 hopping integral $\, t\sb{ij} \,$ we will have the
 electron-hole symmetry and a same $T\sb{c}$.
 In the reduced $p$-$d$ model~(\ref{eq:H}),
the band width for the singlet subband (proportional to $\,
t\sp{22}\sb{ij}\, $) and the one-hole subband (proportional to $\,
t\sp{11}\sb{ij}\, $)  and the corresponding coupling constants are
different that results in different $T\sb{c}(\delta)$ curves as
observed in experiment.~\cite{Schon01}
\par
A substantial proof of the analytical results given above is
provided by a numerical study of the original gap
equation~(\ref{gap22}) using a direct summation in $\bf k$-space.
The numerical solution of the gap equation~(\ref{gap22}) was
performed  under the condition of no double occupation at each
lattice site $i$  by a quantum state $\,|in\rangle\,$ in the
upper, $\, n= 2, \, $ or lower, $\, n=\sigma ,\,$  Hubbard
subbands:
\begin{equation}
  \langle X\sb{i}\sp{\sigma 2} X\sb{i}\sp{\bar\sigma 2} \rangle =
  \langle X\sb{i}\sp{0\bar\sigma} X\sb{i}\sp{0\sigma} \rangle = 0,
\label{eq:constr}
\end{equation}
which automatically follows from the Hubbard operator
multiplication rules. The condition is  satisfied by the order
parameter with ${d}\sb{x\sp{2}-y\sp{2}}$ symmetry for a square
lattice because the anomalous correlation functions $\,\langle
X\sb{\bf k}\sp{\sigma 2} X\sb{-\bf k}\sp{\bar\sigma 2} \rangle \,$
or $\,\langle X\sb{\bf k}\sp{0\bar\sigma} X\sb{-\bf k}\sp{0\sigma}
 \rangle \,$ having the $d\sb{x\sp{2}-y\sp{2}}$-wave symmetry
change the sign under the permutation of the components $k\sb{x}$
and $k\sb{y}$  that results in vanishing of the sum over ${\bf k}$
in the equation: $\,
  \langle X\sb{i}\sp{\sigma 2} X\sb{i}\sp{\bar\sigma 2} \rangle =
({1}/{N}) \sum\sb{\bf k}\,\langle X\sb{\bf k}\sp{\sigma 2}
X\sb{-\bf k}\sp{\bar\sigma 2} \rangle =0 \,.$
\par
 The discretization of the Fredholm integral equation~(\ref{gap22})
results in an eigenvalue problem for the gap parameter
$\bar\Phi\sp{22}\sb{\sigma}({\bf q})$ of the singlet subband. The
temperature at which the largest eigenvalue equals one, while the
corresponding eigenvector shows ${d}\sb{x\sp{2}-y\sp{2}}$
symmetry, will be taken for the critical temperature $T\sb{c}$. We
note that the eigenvalues of the integral equation are discrete
(since the kernel is compact) and real (since the kernel can be
made symmetric).
\par
  The numerical results reported below have been derived under the
following parameter values: $\,\Delta\sb{pd} = 2t\sb{pd} = 3$~eV,
$\, t\sb{eff} \simeq K\sb{22}2\nu\sb{1} t\sb{pd} \simeq 0.14
t\sb{pd} \simeq 0.2$~eV.  For the exchange interaction, we assumed
the value $\, J = 0.4 t\sb{eff},\,$ usually considered in the
$t$-$J$ model. In the model for the static spin
susceptibility~(\ref{eq:oimchi}) the antiferromagnetic correlation
length was equated to the typical value $\xi=3$ which was kept
independent of $\delta$ and for  the cut-off frequency we took
$\omega_s = 0.15$~eV.
\par
\begin{figure}[t]
\centering \epsfig{file=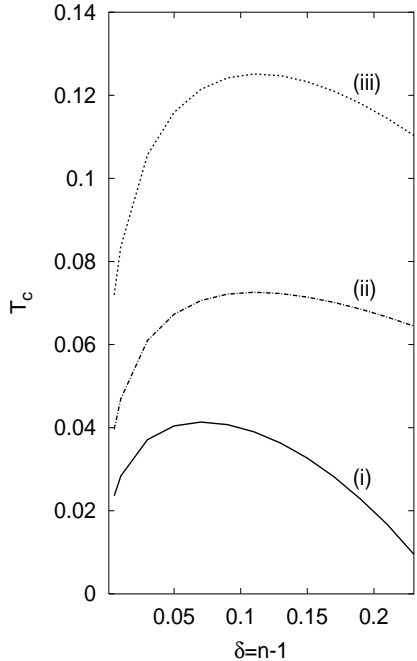,height=90mm,width=60mm}
\vspace*{1mm} \caption{ $T_{c}(\delta)\, $ (in  units of
$\,t_{eff})$ for:
  (i)~kinematic interaction (solid line),
 (ii)~exchange interaction  (dashed line),
 (iii)~for  both contributions (dotted line)}
 \label{fig-tc}
\end{figure}

Fig.~\ref{fig-tc} shows the dependence of the critical
temperature $T\sb{c}$ (in $t\sb{eff}$  units) on the doping
parameter $\delta $.
  The kinematic interaction alone results in a lower $T\sb{c}$ (the solid line
in Fig.~\ref{fig-tc}) as compared to the exchange interaction (the dashed
line). The highest $T\sb{c}$ (dotted line) is obtained when both interactions
are included in Eq.~(\ref{gap22}).
  The maximum values of $T\sb{c}$ are quite high in all cases.
They vary from $\,T\sb{c}\sp{max} \simeq 0.12 t\sb{eff} \simeq
270$~K at optimum doping $\, \delta\sb{opt} \simeq 0.13\,$ in the
highest curve to $\,T\sb{c}\sp{max} \simeq 0.04 t\sb{eff} \simeq
90$~K at $\, \delta\sb{opt} \simeq 0.07\,$ in the lowest curve.
There is a reasonable agreement of these  values with the crude
analytical estimates~(\ref{tcJ}),~(\ref{tcsf}). It is important to
note that the WCA results overestimate $T\sb{c}$ because of the
neglect of the  inelastic scattering which strongly suppresses
$T\sb{c}$ as was proved for the $t$-$J$ model.~\cite{Plakida99} At
smaller antiferromagnetic correlation lengths, the numerical
solution suggested a strong decrease of the contribution of the
kinematic interaction to $T\sb{c}$. This feature stems from the
decrease of the parameter $\chi\sb{0}(\xi)$ which controls the
magnitude of the electron-electron coupling induced by the
spin-fluctuations. The parabolic behavior of $T\sb{c}$ with
$\delta$, showing a maximum at  optimum doping $\, \delta\sb{opt}
\simeq 0.13\,$, is similar to the experimental data. This behavior
originates from a strong dependence of the density of states on
the doping, with a marked peak at the optimum value $\,
\delta\sb{opt}\,$, as shown in Fig.~2 in Ref.~\cite{Plakida95a}
for the  Hubbard model~(\ref{eq:H}). At doping values   $\delta \to 0,
\,$ the superconducting $T\sb{c}$ vanishes. However, at low doping a
phase transition to AFM state should occur and therefore we do not
show $T\sb{c}(\delta)$ for
 $\,\delta \to 0 \,$. Moreover, at low doping a pseudogap
formation should be taken into account that further suppresses
$T\sb{c}$.
\par
\begin{figure}[t]
\centering \epsfig{file=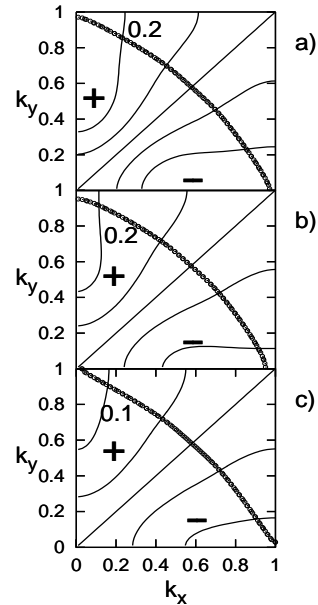,height=80mm,width=40mm}
\vspace*{1mm} \caption{Wave-vector  dependence of the gap function
$\bar\Phi\sp{22}({\bf k})$  over the first quadrant of the BZ  at
optimum doping ($\delta=0.13$) for temperatures:
 $\,T=0\,$~(a), $\,0.5T\sb{c}\,$~(b), and
 $\,0.9T\sb{c}\,\,$~(c). The circles plot the Fermi surface.
 The $+/-$ denote gap signs inside octants.
 The numbers in graphs show the maximum isoline value.}
 \label{fig-gap}
\end{figure}

Further insight is obtained from the analysis of the {\bf
k}-dependence of the gap $\bar\Phi\sp{22}(\bf k)$ inside the BZ
 at   different  temperatures. In Fig.~\ref{fig-gap} a typical
 wave-vector dependence of the gap is shown in the first quadrant
 of the BZ, $\,(0\leq k\sb{x},~k\sb{y}\leq 1,\,$ in $\pi/a$ units),
 at optimum doping ($\delta=0.13$) at three temperatures,
 $\,T=0 \,$~(a), $\,T=0.5T\sb{c} \,$~(b) and $T=0.9T\sb{c} \,$~(c).
The occurrence of a superconducting state with
${d}\sb{x\sp{2}-y\sp{2}}$ symmetry is observed, with a ${\bf
q}$-dependence pattern of the gap shown by isolines. This
dependence is much more complicated than that depicted by the
simple analytical model~(\ref{g1}).
\par
 Consideration of the
kinematic interaction term alone in Eq.(\ref{gap22}) and the
subsequent numerical solution at the specific optimum doping value
$\, \delta=0.07,\,$ resulted in gap maxima inside the BZ, while
inclusion of the exchange interaction as well and numerical
solution at the higher optimum doping value $\, \delta=0.13 \,$
resulted in the shift of the maxima towards the $(\pi,0)$-type
points of the BZ, which are close to the FS, as shown in
Fig.~\ref{fig-gap}. The particular behavior of the
spin-fluctuation pairing is explained by vanishing of the
kinematic interaction along the lines $\,|k_x|+|k_y| = \pi \,$,
close to the large FS, as discussed above. These results  show
that the inclusion of the exchange term is essential for the
achievement of experimentally observed in cuprates wave-vector
dependence of the gap and an optimum doping values. An additional
experimental evidence for AFM exchange pairing mechanism in
cuprates comes from the $T\sb{c}$ dependence on the lattice
constants or an external pressure. While in most electron-phonon
superconductors $T\sb{c}$ decreases with an increasing applied
pressure, in cuprates $T\sb{c}$ (at optimum doping) increases with
pressure.~\cite{Schilling01} As was shown
recently~\cite{Plakida01}, the AFM exchange pairing model for
$T\sb{c}$~(\ref{tcJ}) describes quite well the experimentally
observed $T\sb{c}$ dependence on the lattice constants in the
CuO$\sb{2}$-plane in mercury cuprates.
\par
Therefore, we may conclude that the most important pairing
interaction in the strong correlation limit of the Hubbard
model~(\ref{eq:H}) is the exchange interaction, while the
spin-fluctuation coupling mediated by the kinematic interaction
results only in a moderate enhancement of $T\sb{c}$. The same
result has been obtained in studies of the $t$-$J$ model beyond
the WCA in Ref.~\cite{Plakida99}. However, the
superconducting
exchange pairing in the $t$-$J$ model is realized already in the
simplest MFA  for the nearest neighbor particles, while in the
Hubbard model a more complicated three-particle anomalous
correlation functions, Eq.~(\ref{eq:hac}), are induced by a
dynamical  second order interband hopping responsible for the
exchange interaction. This observation may explain why  numerical
simulations for the $t$-$J$ model~\cite{Sorella01} usually
predict a much stronger pairing tendency than for the original
Hubbard model~\cite{Huang01} since in the latter  more complicated
dynamical processes should be accounted for.

\section{Conclusions}
\label{sec:concl}
A microscopical theory of the superconducting pairing has been
proposed within the reduced two-band $p$-$d$ Hubbard
model~(\ref{eq:H}) with two microscopical parameters only, the
$p$-$d$ hybridization parameter $t\sb{pd}$ and the charge-transfer
gap $\Delta\sb{pd}$. By applying the projection technique
 for the two-time GF in terms of the Hubbard
operators, we obtained the Dyson equation for the $4 \times 4$
matrix GF~(\ref{eq:Dyson}) with the corresponding matrix
self-energy~(\ref{eq:self-enir-qo}).
\par
 In the zero-order approximation for  GF, Eq.~(\ref{G22a}), we have
proved that the $d$-wave  superconducting  pairing of conventional
hole (electron) pairs in one Hubbard subband occurs that is
mediated by the conventional exchange interaction $\, J\sb{ij} =
4 (t\sb{ij})\sp{2} / \Delta$  caused by the interband hopping as
in the $t$-$J$ model. Contrary to
Refs.~\cite{Beenen95,Stanescu00}, the anomalous correlation
function in MFA $\, \langle X\sb{i}\sp{02} N\sb{j} \rangle
  = \langle c\sb{i\downarrow}c\sb{i\uparrow} N\sb{j}\rangle \,$
was calculated {\it without any decoupling approximation} (see
Sec.~\ref{sec:dpairing}). We have also proved that  retardation
effects in the exchange interaction are negligible that results
in pairing of all electrons (holes) in the conduction band and a
high-$T_c$ proportional to the Fermi energy, Eq.~(\ref{tcJ}).
\par
Spin-fluctuation  pairing  was studied in the second order of the
intraband hopping within the self-consistent Born approximation
for the self-energy~(\ref{eq:phi-qo}). The observed
spin-fluctuation $d$-wave pairing in the Hubbard model is mediated
by the kinematic interaction which vanishes inside the BZ along
the lines $\, |k\sb{x}| + |k\sb{y}|=\pi \,$ and produces pairing
only in a narrow energy shell of the order of spin-fluctuation
energy close to the FS: $\,\omega_s \ll E_F \,$. These specific
properties of the interaction  results in a lower $ T_c $ mediated
by spin-fluctuations, Eq.~(\ref{tcsf}), in comparison with the
exchange interaction, Eq.~(\ref{tcJ}). Numerical solution of the
gap equation~(\ref{gap22})  confirms the analytical estimations
for the superconducting $T_c$ and reproduces experimentally
observed $T_c(\delta)$ dependence with a maximum at the optimum
doping $\delta\sb{opt} \simeq 0.13\,$,  Fig.~\ref{fig-tc}.
 The $d$-wave symmetry of the gap function with a
complicated ${\bf k}$- dependence, Fig.~\ref{fig-gap}, is also
proved by the numerical solution.
\par
It is to be stressed that, within the Hubbard model in the limit
of strong correlations, the electron-electron coupling  induced by
the exchange and kinematic interactions are caused by  the
non-fermionic commutation relations for the Hubbard operators, and
therefore no additional fitting parameters  are needed to
accommodate that interactions. These mechanisms of superconducting
pairing are absent in the fermionic models as was pointed out by
Anderson.~\cite{Anderson97} However, they  are different: while
the AFM exchange pairing originates in the lowering of kinetic
energy of electron (hole) pairs caused by the interband hopping as
suggested by Anderson~\cite{Anderson87}, the spin-fluctuation
pairing is of the conventional nature: the pairing is due to
attraction in the $d$-wave channel caused by the spin-fluctuation
exchange.
\par
The weak point of the argument is the derivation of the reported
results in the weak coupling approximation for the superconducting
equations. To substantiate the present results, more rigorous
self-consistent   numerical solution of the strong coupling Dyson
 equations in the $({\bf q}, \omega)$-space, as done
for the $t$-$J$ model, \cite{Plakida99} should be elaborated. In
this connection  we can mention
papers~\cite{Maier00,Lichtenstein00} where superconducting pairing
in the Hubbard model was investigated within the non-crossing
approximation for the self-energy in the 4-cluster model in the
dynamical mean field theory.

\acknowledgments
We thank P.~Horsch, F.~Mancini, and V.S.~Oudovenko for stimulating
discussions. One of the authors (N.P.) is grateful to
Prof.~P.~Fulde for the hospitality extended to him during his stay
at MPIPKS, where a part of the present work was done.
Two authors (S.A. and Gh.A.) acknowledge partial financial support by
Romanian MER (grant 7038GR).
\appendix

\section{ Frequency matrix}
To  calculate  the frequency matrix~(\ref{eq:freq})  one  needs
to consider the equations of motion of the Hubbard operators:
\begin{eqnarray}
  Z\sb{i}\sp{\sigma 2} & = & [X\sb{i}\sp{\sigma 2}, H] =
   (E\sb{1} + \Delta) X\sb{i}\sp{\sigma2}
\nonumber \\
 &+&  \sum\sb{l\ne i,\sigma '}\! \left( t\sb{il}\sp{22}
    B\sb{i\sigma\sigma '}\sp{22} X\sb{l}\sp{\sigma ' 2} -
   2 \sigma t\sp{21}\sb{il} B\sb{i\sigma\sigma '}\sp{21}
    X\sb{l}\sp{0\bar\sigma '} \right)
\nonumber \\
  & - & \sum\sb{l\ne i} X\sb{i}\sp{02} \left( t\sp{11}\sb{il}
    X\sb{l}\sp{\sigma0} + 2 \sigma t\sp{21}\sb{il}
    X\sb{l}\sp{2 \bar\sigma} \right),
\label{sec:A1}\\
  Z\sb{i}\sp{0 \bar\sigma } & = & [X\sb{i}\sp{0\bar\sigma }, H] =
   E\sb{1} X\sb{i}\sp{0\bar\sigma}
\nonumber \\
  & + & \sum\sb{l\ne i,\sigma '} \left( t\sb{il}\sp{11}
    B\sb{i\sigma\sigma '}\sp{11} X\sb{l}\sp{0 \bar\sigma '}\!\! -
   2 \sigma t\sp{12}\sb{il} B\sb{i\sigma\sigma '}\sp{12}
    X\sb{l}\sp{\sigma ' 2} \right)
\nonumber \\
  & - & \sum\sb{l\ne i} X\sb{i}\sp{02} \left( t\sp{22}\sb{il}
    X\sb{l}\sp{2\bar\sigma} + 2 \sigma t\sp{12}\sb{il}
    X\sb{l}\sp{\sigma0} \right),
\label{sec:A2}\\
  Z\sb{i}\sp{2 \bar \sigma} & = & - \left(
   Z\sb{i}\sp{\bar\sigma 2}\right)\sp{\dagger},\quad Z\sb{i}\sp{\sigma 0} =
  - \left( Z\sb{i}\sp{0 \sigma} \right)\sp{\dagger}.
\label{sec:A3}
\end{eqnarray}
Here, $B\sb{i\sigma\sigma'}\sp{\alpha\beta}$ are Bose-like operators
describing the number (charge) and spin fluctuations:
\begin{eqnarray}
  B\sb{i\sigma\sigma'}\sp{22} & = & (X\sb{i}\sp{22} +
   X\sb{i}\sp{\sigma\sigma}) \delta\sb{\sigma'\sigma} +
   X\sb{i}\sp{\sigma\bar\sigma} \delta\sb{\sigma'\bar\sigma}
\nonumber \\
  & = & (\frac{1}{2} N\sb{i} + S\sb{i}\sp{z}) \delta\sb{\sigma'\sigma} +
    S\sb{i}\sp{\sigma}\delta\sb{\sigma'\bar\sigma},
  \label{sec:A4}\\
  B\sb{i\sigma\sigma'}\sp{21} & = & (\frac{1}{2} N\sb{i} +
   S\sb{i}\sp{z}) \delta\sb{\sigma'\sigma} -
   S\sb{i}\sp{\sigma} \delta\sb{\sigma'\bar\sigma},
  \label{sec:A5}\\
  B\sb{i\sigma\sigma'}\sp{11} & = & \delta\sb{\sigma'\sigma} -
   B\sb{i\sigma\sigma'}\sp{21}, \quad
  B\sb{i\sigma\sigma'}\sp{12} = \delta\sb{\sigma'\sigma} -
   B\sb{i\sigma\sigma'}\sp{22} ,
\label{sec:A6}
\end{eqnarray}
where we have used  the completeness relation~(\ref{eq:no2oc}),
and the definition of the number operator $N\sb{i}$,
Eq.~(\ref{def-n}) and the spin operators $ \quad S\sb{i}\sp{z}=
\sum\sb{\sigma}\sigma X\sb{i}\sp{\sigma\sigma}, \quad
  S\sb{i}\sp{\sigma} = X\sb{i}\sp{\sigma\bar\sigma} \;. \quad  $
 Performing necessary commutations of the operators $\,
Z\sb{i}\sp{\sigma 2}\,$ and $\,  Z\sb{i}\sp{0 \bar\sigma }  \, $
in the matrix $\tilde {\cal A}\sb{ij\sigma}$,
Eq.~(\ref{eq:Aij}),  we get  for  the quantities $a\sp{\alpha
\beta}\sb{\sigma}$  in Eq.~(\ref{omega}) the following equations
\begin{eqnarray}
  a\sp{22}\sb{\sigma} & = & \sum\sb{m\neq i} V\sb{im} \left(
   K\sb{22} \bigl< X\sb{i}\sp{2\bar \sigma} X\sb{m}\sp{\bar \sigma 2} \bigr> -
   K\sb{11} \bigl< X\sb{m}\sp{\sigma 0} X\sb{i}\sp{0\sigma} \bigr> \right),
\label{sec:A7}\\
  a\sp{21}\sb{\sigma} & = & - \sum\sb{m\neq i} V\sb{im} \left(
   K\sb{22} \bigl< X\sb{i}\sp{\sigma 0} X\sb{m}\sp{\bar \sigma 2}\bigr> +
   K\sb{11} \bigl< X\sb{i}\sp{\bar\sigma 2} X\sb{m}\sp{\sigma 0} \bigr> \right)
\nonumber \\
  & - & 2\sigma \sum\sb{m\neq i} V\sb{im} K\sb{12} \left(
   \bigl< X\sb{i}\sp{\sigma 0} X\sb{m}\sp{0\sigma} \bigr> -
   \bigl< X\sb{m}\sp{2\bar\sigma} X\sb{i}\sp{\bar\sigma 2} \bigr> \right).
\label{sec:A8}
\end{eqnarray}
For the renormalized hopping parameters $K\sb{ij\sigma}\sp{\alpha \beta}$
in the matrix~(\ref{omega}) analogous calculations give the results
\begin{eqnarray}
  K\sb{ij\sigma}\sp{22} & = & K\sb{22} \chi\sb{ij}\sp{cs} -
   K\sb{11} \bigl< X\sb{i}\sp{02} X\sb{j}\sp{20} \bigr>,
\label{sec:A9}\\
  K\sb{ij\sigma}\sp{11} & = & K\sb{11} \left( \chi\sb{ij}\sp{cs} + 1 -
   n \right) - K\sb{22} \bigl< X\sb{i}\sp{02} X\sb{j}\sp{20} \bigr>,
  \label{sec:A10}\\
  K\sb{ij\sigma}\sp{21} & = & 2\sigma K\sb{12} \left( \chi\sb{ij}\sp{cs} -
   \frac{1}{2} n - \bigl< X\sb{i}\sp{02} X\sb{j}\sp{20} \bigr> \right),
\label{sec:A11}
\end{eqnarray}
where $\chi\sb{ij}\sp{cs}$ stands for the static charge and spin
correlation functions
\begin{equation}
 \chi\sb{ij}\sp{cs} = \frac{1}{4} \langle N\sb{i} N\sb{j}
\rangle +
   \langle {\bf S}\sb{i} {\bf S}\sb{j} \rangle .
\label{sec:A12}
\end{equation}
In Eq.~(\ref{delta}), the site independent anomalous correlation functions
are given respectively by
\begin{eqnarray}
  b\sp{22}\sb{\sigma} & = & \sum\sb{m\neq i} V\sb{im} \bigl\{ K\sb{22}
   \left( \bigl< X\sb{i}\sp{\bar\sigma2} X\sb{m}\sp{\sigma 2} \bigr> -
   \bigl< X\sb{i}\sp{\sigma2} X\sb{m}\sp{\bar\sigma 2} \bigr> \right)
\nonumber \\
  & - & 2 \sigma K\sb{12} \left( \bigl< X\sb{i}\sp{\sigma 2}
    X\sb{m}\sp{0\sigma} \bigr> + \bigl< X\sb{i}\sp{\bar\sigma 2}
    X\sb{m}\sp{0\bar\sigma} \bigr> \right) \bigr\},
 \label{sec:A13}\\
  b\sp{11}\sb{\sigma} & = & -\sum\sb{m\neq i} V\sb{im} \bigl\{ K\sb{11}
   \left( \bigl< X\sb{i}\sp{0\sigma} X\sb{m}\sp{0\bar\sigma} \bigr> -
   \bigl< X\sb{i}\sp{0\bar\sigma} X\sb{m}\sp{0\sigma} \bigr> \right)
\nonumber \\
  & - & 2 \sigma K\sb{12} \left( \bigl< X\sb{i}\sp{0\sigma}
    X\sb{m}\sp{\sigma2} \bigr> + \bigl< X\sb{i}\sp{0\bar\sigma}
    X\sb{m}\sp{\bar\sigma2} \bigr> \right) \bigr\},
\label{sec:A14}\\
  b\sp{21}\sb{\sigma} & = & \sum\sb{m\neq i} V\sb{im} \bigl\{ K\sb{22}
   \left( \bigl< X\sb{i}\sp{0\sigma} X\sb{m}\sp{\sigma2} \bigr> +
   \bigl< X\sb{i}\sp{0\bar\sigma} X\sb{m}\sp{\bar\sigma2} \bigr> \right)
\nonumber \\
  & - & 2 \sigma K\sb{12} \left( \bigl< X\sb{i}\sp{0\sigma}
    X\sb{m}\sp{0\bar\sigma} \bigr> - \bigl< X\sb{i}\sp{0\bar\sigma}
    X\sb{m}\sp{0\sigma} \bigr> \right) \bigr\}.
\label{sec:A15}
\end{eqnarray}
The site-dependent anomalous correlation functions are convenient to
write as follows
\begin{eqnarray}
  L\sb{ij\sigma}\sp{22} & = & - 2 \sigma K\sb{21}
   \bigl< X\sb{i}\sp{02} N\sb{j} \bigr>,
\label{sec:A16}   \\
  L\sb{ij\sigma}\sp{11} & = & - 2 \sigma  K\sb{21}
   \bigl<(2 - N\sb{j})  X\sb{i}\sp{02} \bigr>,
 \label{sec:A17}  \\
   L\sb{ij\sigma}\sp{21} & = & \frac{1}{2 }
  \bigl( K\sb{22}\bigl< X\sb{i}\sp{02} N\sb{j} \bigr> -
    K\sb{11} \bigl<(2 - N\sb{j})  X\sb{i}\sp{02} \bigr> \bigr) .
\label{sec:A18}
\end{eqnarray}
According to Eq.~(\ref{eq:x2da}), the anomalous correlation
functions describe the pairing at one lattice site but in
different subbands: $\bigl< X\sb{i}\sp{02} N\sb{j}\bigr> = \bigl<
X\sb{i}\sp{0\downarrow} X\sb{i}\sp{\downarrow 2}  N\sb{j}\bigr> =
\bigl< c\sb{i\downarrow}c\sb{i\uparrow}  N\sb{j} \bigr>$.

\section{ Self-energy}
 The starting point for the calculation of self-energy
corrections is the $({\bf r}, t)$-representation of the
self-energy~(\ref{eq:self-enir-qo}):
\begin{equation}
  \tilde \Sigma\sb{ij\sigma}(t-t') = \tilde\chi\sp{-1}
   \langle\!\langle
    \hat Z\sb{i\sigma}\sp{(ir)}(t) \!\mid\!
    \hat Z\sb{j\sigma}\sp{(ir)\dagger}(t')
   \rangle\!\rangle\sp{(prop)}\tilde \chi\sp{-1}.
\label{sec:B1}
\end{equation}
The operators $Z\sp{(ir)}$ are obtained from
Eqs.~(\ref{sec:A1})-(\ref{sec:A3}) according to definition,
Eq.~(\ref{eq:irred}). Neglecting the terms which contain
$X\sb{i}\sp{02}$ operators, whose contribution to the low energy
dynamics of the system is assumed to be small (since it involves
large charge fluctuations), we get:
\begin{eqnarray} \
  Z\sb{i,\sigma 2}\sp{(ir)} &=& \sum\sb{l\ne i,\sigma'} \bigl(
   t\sb{il}\sp{22} \delta B\sb{i\sigma\sigma'}\sp{22} X\sb{l}\sp{\sigma'2} -
   2 \sigma t\sp{21}\sb{il} \delta B\sb{i\sigma\sigma'}\sp{21}
    X\sb{l}\sp{0 \bar\sigma'} \bigr),
 \label{sec:B2}\\
Z\sb{i,0 \bar\sigma}\sp{(ir)} &=& \sum\sb{l\ne i,\sigma'} \bigl(
t\sb{il}\sp{11} \delta B\sb{i\sigma\sigma'}\sp{11}
    X\sb{l}\sp{0 \bar\sigma '}\!\! -
   2 \sigma t\sp{12}\sb{il} \delta B\sb{i\sigma\sigma'}\sp{12}
    X\sb{l}\sp{\sigma ' 2} \bigr) . \!\!
\label{sec:B3}
\end{eqnarray}
 Here, $\, \delta
B\sb{i\sigma\sigma'}\sp{\alpha\beta} =
B\sb{i\sigma\sigma'}\sp{\alpha\beta} - \langle
B\sb{i\sigma\sigma'}\sp{\alpha\beta} \rangle \,$ are Bose-like
operators describing charge and spin fluctuations as  follows
 from Eqs.~(\ref{sec:A4})-(\ref{sec:A6}).
\par
The many-particle GF in Eq.~(\ref{sec:B1}) associated to the
irreducible operators, Eqs.~(\ref{sec:B2}),~(\ref{sec:B3}), give
the following $2\times 2$ matrices $\hat M$ and $\hat\Phi$ for the
 normal and anomalous contributions to the self-energy,
respectively:
\begin{eqnarray}
  \hat M\sb{ij\sigma}(\omega) &\! =\! &
  \left( \begin{array}{cc}
    M\sb{ij\sigma}\sp{22}(\omega) & M\sb{ij\sigma}\sp{21}(\omega)\\
    M\sb{ij\sigma}\sp{12}(\omega) & M\sb{ij\sigma}\sp{11}(\omega)
  \end{array} \right)
\nonumber \\
  &\! = &\! \left\langle\!\!\left\langle\!\!
  \left(\!\! \begin{array}{c}
    \left(Z\sb{i}\sp{\sigma 2}\right)\sp{(ir)}\!\! \\
    \left(Z\sb{i}\sp{0\bar\sigma}\right)\sp{(ir)}\!\!
  \end{array} \right)\!\!  \Bigg |\!
  \left(\!\! \begin{array}{cc}
    \left(Z\sb{j}\sp{2 \sigma}\right)\sp{(ir)}\!\! &\!\!
    \left(Z\sb{j}\sp{\bar\sigma 0}\right)\sp{(ir)}\!\!
  \end{array} \right)\!\! \right\rangle\!\!\right\rangle \sb{\!\!\omega}\, ,
\label{sec:B4}\\
  \hat \Phi\sb{ij\sigma}(\omega) &\! =\! &
  \left( \begin{array}{cc}
    \Phi\sb{ij\sigma}\sp{22}(\omega) & \Phi\sb{ij\sigma}\sp{21}(\omega)\\
    \Phi\sb{ij\sigma}\sp{12}(\omega) & \Phi\sb{ij\sigma}\sp{11}(\omega)
  \end{array} \right)
\nonumber \\
  &\! = &\! \left\langle\!\!\left\langle\!\!
  \left(\!\! \begin{array}{c}
    \left(Z\sb{i}\sp{\sigma 2}\right)\sp{(ir)}\!\! \\
    \left(Z\sb{i}\sp{0\bar\sigma}\right)\sp{(ir)}\!\!
  \end{array} \right)\!\!  \Bigg |\!
  \left(\!\! \begin{array}{cc}
    \left(Z\sb{j}\sp{\bar\sigma 2}\right)\sp{(ir)}\!\! &\!\!
    \left(Z\sb{j}\sp{0\sigma}\right)\sp{(ir)}\!\!
  \end{array} \right)\!\! \right\rangle\!\!\right\rangle \sb{\!\!\omega}\, .
\label{sec:B5}
\end{eqnarray}
In the diagonal approximations for the GF  the self-energy
 matrix elements  $\hat M\sb{ij\sigma}(\omega)$~(\ref{sec:B4}) and
$\hat \Phi\sb{ij\sigma}(\omega)$~(\ref{sec:B5}) within the
SCBA~(\ref{eq:decoupl-ro}) in the $({\bf q},
\omega)$-representation are given by the equations:
\begin{eqnarray}
  \hat M\sb{\sigma}({\bf q},\omega) = \frac{1}{N} \sum\sb{\bf k}
   \int\limits\sb{-\infty}\sp{+\infty} \!\!{\rm d}\omega\sb{1}
    K\sp{(+)}(\omega,\omega\sb{1}|{\bf k},{\bf q - k})
\nonumber \\
  \times \; \left\{ - \frac{1}{\pi} \mbox{Im} \left[ \hat P\sb{2}\sp{(+)}
   G\sb{\sigma}\sp{22}({\bf k},\omega\sb{1}) + \hat P\sb{1}\sp{(+)}
   G\sb{\sigma}\sp{11}({\bf k},\omega\sb{1}) \right] \right\} \; ,
\label{sec:B6} \\
  \hat\Phi\sb{\sigma}({\bf q},\omega) = \frac{1}{N} \sum\sb{{\bf k}}
   \int\limits\sb{-\infty}\sp{+\infty} \!\!{\rm d}\omega\sb{1}
    K\sp{(-)}(\omega,\omega\sb{1}|{\bf k},{\bf q - k})
\nonumber \\
  \times \; \left\{ - \frac{1}{\pi} \mbox{Im} \left[ \hat P\sb{2}\sp{(-)}
   F\sb{\sigma}\sp{22}({\bf k},\omega\sb{1}) - \hat P\sb{1}\sp{(-)}
   F\sb{\sigma }\sp{11}({\bf k},\omega\sb{1}) \right] \right\} \, ,
\label{sec:B7}
\end{eqnarray}
where
\begin{eqnarray}
  \hat P\sb{2}\sp{(\pm)} &\! =\! &
    \left( \begin{array}{lll}
      K\sb{22}\sp{2} & & \pm 2 \sigma K\sb{21} K\sb{22}\\
      2 \sigma K\sb{21} K\sb{22} & & \pm K\sb{21}\sp{2}
    \end{array} \right) \, ,
\label{sec:B8} \\
  \hat P\sb{1}\sp{(\pm)} &\! =\! &
    \left( \begin{array}{lll}
      K\sb{21}\sp{2} & & \pm 2 \sigma K\sb{21} K\sb{11}\\
      2 \sigma K\sb{21} K\sb{11} & & \pm K\sb{11}\sp{2}
    \end{array} \right) .
\label{sec:B9}
\end{eqnarray}
The diagonal matrix elements $\, M\sb{\sigma}\sp{22}({\bf q},\omega), \;
 \Phi\sb{\sigma}\sp{22}({\bf q},\omega) ,\;$ for the singlet subband are given
 by Eqs.~(\ref{eq:m-qo}),~(\ref{eq:phi-qo}), while for the one-hole subband,
 $\, M\sb{\sigma}\sp{11}({\bf q},\omega), \;
 \Phi\sb{\sigma}\sp{11}({\bf q},\omega) ,\;$  can be easily obtained from
 the matrices~(\ref{sec:B6}),~(\ref{sec:B7}).


%
\end{document}